\documentclass[aps,pra,reprint,amsmath,amssymb,superscriptaddress,twocolumn,longbibliography,nofootinbib,notitlepage]{revtex4-2}
\usepackage{mathtools}
\usepackage{braket}
\usepackage{siunitx}
\usepackage[hidelinks]{hyperref}
\usepackage{svg}
\usepackage{caption}
\usepackage{subcaption}
\usepackage{dsfont}
\usepackage{placeins}
\usepackage{bbm}
\usepackage{quantikz}
\usepackage{tikz}

\newcommand\tab[1][0.3cm]{\hspace*{#1}}

\newcommand{\qft}{\hat{\textrm{QFT}}}
\newcommand{\mycomment}[1]{}

\UseRawInputEncoding

\usepackage{xcolor}

\begin{document}
\title{Quantum variational solving of nonlinear and multi-dimensional \\ partial differential equations}
\author{Abhijat~Sarma}
\email{as3232@cornell.edu}
\affiliation{Department of Physics, Cornell University, Ithaca, NY 14853, USA}
\author{Thomas~W.~Watts}
\affiliation{School of Applied and Engineering Physics, Cornell University, Ithaca, NY 14853, USA}
\author{Mudassir~Moosa}
\affiliation{Department of Physics, Cornell University, Ithaca, NY 14853, USA}
\affiliation{Department of Physics and Astronomy, Purdue University, West Lafayette, IN 47907, USA}
\author{Yilian~Liu}
\affiliation{School of Applied and Engineering Physics, Cornell University, Ithaca, NY 14853, USA}
\author{Peter~L.~McMahon}
\email{pmcmahon@cornell.edu}
\affiliation{School of Applied and Engineering Physics, Cornell University, Ithaca, NY 14853, USA}

\begin{abstract}
A variational quantum algorithm for numerically solving partial differential equations (PDEs) on a quantum computer was proposed by Lubasch et al. \cite{lubasch}. In this paper, we generalize the method introduced by Lubasch et al. to cover a broader class of nonlinear PDEs as well as multidimensional PDEs, and study the performance of the variational quantum algorithm on several example equations. Specifically, we show via numerical simulations that the algorithm can solve instances of the Single-Asset Black--Scholes equation with a nontrivial nonlinear volatility model, the Double-Asset Black--Scholes equation, the Buckmaster equation, and the deterministic Kardar--Parisi--Zhang equation. Our simulations used up to $n=12$ ansatz qubits, computing PDE solutions with $2^n$ grid points. We also performed proof-of-concept demonstrations with a trapped-ion quantum processor from IonQ \cite{IONQ}, showing accurate computation of two representative expectation values needed for the calculation of a single timestep of the nonlinear Black--Scholes equation. Through our classical simulations and demonstrations on quantum hardware, we have identified---and we discuss---several open challenges for using quantum variational methods to solve PDEs in a regime with a large number ($\gg 2^{20}$) of grid points, but also a practical number of gates per circuit and circuit shots.
\end{abstract}

\maketitle

\section{Introduction}
\label{sec:introduction}
Quantum computing is an alternative paradigm for computation that, for some applications, may enable an exponential advantage in space (memory) and/or time versus classical approaches. One recently proposed application of quantum computing is to solving partial differential equations (PDEs) via a variational approach \cite{lubasch}. The general approach in variational quantum algorithms \cite{Cerezo2020-1} is as follows. A proposed solution to a problem is represented as a quantum state that is prepared using a low-depth parameterized quantum circuit, often called the ansatz circuit. A cost function, which ideally should be easily computable given a proposed solution state, quantifies the quality of the state as a solution to the problem being solved. An optimizer running on a classical computer is used to update the parameters of the ansatz circuit with the goal of minimizing the cost function. Variational quantum algorithms tend to have lower qubit-count and circuit-depth requirements than more traditional quantum algorithms \cite{nielsen2010quantum} and partially for this reason have attracted attention as candidates to run on noisy intermediate-scale quantum (NISQ) processors \cite{preskill2018quantum}.

Nonlinear and multidimensional PDEs are typically intractable analytically and classical numerical methods can be very costly---especially for multidimensional PDEs in which the curse of dimensionality, where costs increase exponentially with the PDE dimension, can arise. Following the publication of the proposal by Lubasch~et~al. \cite{lubasch} to solve PDEs variationally on quantum computers, several studies have explored variational approaches to solving a variety of PDEs \cite{Kyriienko2020, liu2021variational, Mocz2021, Sato2021, garcia2022quantum, YewLeong2022, YewLeong2023, Ewe2022, Rebentrost2023, Amaro2023}.\footnote{There is also a rich and interesting literature (e.g., Refs.~\cite{clader2013preconditioned,cao2013quantum,montanaro2016quantum,Arrazola2019,Childs2021,Jin2022,Amaro2023}) studying how quantum computers could solve PDEs using non-variational approaches, such as via Hamiltonian simulation or using quantum-linear-systems-solving \cite{harrow2009quantum} as a subroutine, which we will not review.} However, not all of these methods are applicable to nonlinear or multidimensional problems, and thus far none have been demonstrated in numerical simulation with more than 6 ansatz qubits (i.e., $2^6 = 64$ grid points) on nonlinear problems. In Ref.~\cite{lubasch}, Lubasch et al. introduced a method to solve linear homogeneous PDEs and a subset of nonlinear PDEs that are first-order in time; they demonstrated their method's applicability to the 1D heat equation and Burger's equation. In Ref.~\cite{Mocz2021}, Mocz and Szasz show how to use Lubasch et al.'s method, together with a subroutine to train a state to represent a nonlinear potential, to solve the Schr\"odinger--Poisson equation. In Ref.~\cite{YewLeong2022}, Yew Leong et al. solve Reaction-Diffusion equations in a similar manner. In this paper, we present a general version of this quantum method for solving PDEs, and demonstrate explicitly how to train intermediate quantum states to represent a large class of nonlinearities and inhomogeneous terms, making the method applicable to most nonlinear and multidimensional PDEs of interest which are first order in time, including the Navier--Stokes equations, Maxwell's equations, the Calabi flow equation, and the Gross--Pitaevski equation, with applications to fluid mechanics, electromagnetism, differential geometry, and Bose-Einstein condensates, respectively. We demonstrate how the method can be used to adapt many explicit and semi-implicit numerical schemes, such as fourth order Runge-Kutta (RK4). We further introduce novel ansaetze well-suited for solving differential equations, and benchmark them with several popular out-of-the-box optimizers. We then apply the method to the solution of a nonlinear Black--Scholes equation (BSE) with 8 ansatz qubits, the 2D linear BSE with 6 ansatz qubits per degree of freedom (giving $6\cdot2=12$ total ansatz qubits), the Buckmaster equation with 5 ansatz qubits, and the deterministic KPZ equation with 5 ansatz qubits, demonstrating applications to options pricing, viscous fluid, and surface growth problems. Finally, we show 2 qubit proof-of-concept results on an 11-qubit trapped-ion device offered by IonQ. The hardware details of the device can be found in Ref.~\cite{IONQ}.


\begin{figure*}
\begin{tikzpicture}
\draw[color=white] (0,4.0)--(0.35,4.0)--(0.35,4.1)--(0.0,4.1)--(0,4.0);
\draw[color=blue!70] (10.35-10.85-0.7-0.5,4.0)--(20.35-3.35-0.5,4)--(20.35-3.35-0.5,-6.8)--(10.35-10.85-0.7-0.5,-6.8)--(10.35-10.85-0.7-0.5,4);
\draw[color=blue!70] (7.5, 4) -- (7.5, -6.8);
\draw[color=blue!70] (7.5, -0.7) -- (16.5, -0.7);

\node at (8.95-10.85-0.7+0.35+4.30-0.5,4-0.5) {\textbf{(a)} \underline{Quantum PDE Solving Framework} };

 \node at (8.75-10.85-0.7+0.35+4.30-0.5, 4-1.5) {\large{$\frac{\partial u}{\partial t} = \hat{O}[u]$}};

 \draw [->] (1.325,2.2) -- (1.325,1.75);

 \node at (8.75-10.85-0.7+0.35+4.30-0.5, 1.5) {Step 1: Read in $u(t_i)$};
 \node at (8.75-10.85-0.7+0.35+4.30-0.5, 1) {\small{$\ket{u(t_i)}\equiv \ket{\tilde{u}}=\tilde{\lambda}_0\hat{U}(\boldsymbol{\tilde{\lambda}})\ket{0}$}};

\draw [->] (1.325,0.7) -- (1.325,0.25);

\node at (9.55-10.85-0.7+0.35+4.30-0.5, -0.15) {Step 2: Decompose $\hat{O}$ into quantum gates (\textbf{(b)})};

\draw [->] (1.325,-0.45) -- (1.325,-1);

\node at (8.95-10.85-0.7+0.35+4.30-0.5, -1.55) {Step 3: Find $u(t+\tau)$ from $u(t)$};
\node at (8.75-10.85-0.7+0.35+4.30-0.5, -2.05) {\small{$\ket{u(t_i+\tau)}\equiv\ket{u}=\lambda_0\hat{U}(\boldsymbol{\lambda})\ket{0}$}};
\node at (9.15-10.85-0.7+0.35+4.30-0.5, -2.55) {\small{$C_u(\lambda_0, \boldsymbol{\lambda})=||(\mathbf{1}-\tau\hat{O})\ket{u} - \ket{\tilde{u}}||^2$}};
\node at (9.03-10.85-0.7+0.35+4.30-0.5, -3.05) {\small{Minimize $C_u(\lambda_0, \boldsymbol{\lambda})$ via classical}};
\node at (9.35-10.85-0.7+0.35+4.30-0.5, -3.48) {\small{optimization loop to find optimal $\ket{u}$}};

\draw (11-10.85-0.7+0.35+4.30-0.2, -1.6) arc (-90:70:.5);

\draw [->] (4.09, -0.63) -- (3.89, -0.65+0.3*0.36);

\node[text width=2.5cm] at (6.1, -1.2) {\tiny{Repeat for desired number of timesteps}};

\draw [->] (1.325,-3.8) -- (1.325,-4.25);
\node at (8.75-10.85-0.7+0.35+4.30-0.5, -4.5) {Step 4: Read Out $u(t_f)$};

\node at (11.2, 3.5) {\textbf{(b)} \underline{Train Intermediate States to Represent $\hat{O}$} };

\node[align=left] at (11.4, 1.3) {\small{for $i$ in $1, ..., N$:} \\
    \tab \small{let} ${\displaystyle \ket{\chi_i} = \theta_0^i \hat{U}(\boldsymbol{\theta}^i)\ket{0}}$ \\
    \tab \small{for $k$ in $0, ..., i-1$:}\\
    \tab \tab \small{let} ${\displaystyle O_i^k = f^k(\hat{A}, \hat{D}_{\tilde{u}}, \hat{D}_{\chi_1}, ..., \hat{D}_{\chi_{i-1}})}$\\
    \tab \small{let} ${\displaystyle C_{\chi_i}(\theta_0^i, \boldsymbol{\theta}^i) = ||\ket{\chi_i} - \hat{O}_i^0\ket{\tilde{u}} - \sum_{k=1}^{i-1} \hat{O}_i^k\ket{\chi_k}||^2}$\\
    \tab \small{Minimize} ${\displaystyle C_{\chi_i}(\theta_0^i, \boldsymbol{\theta}^i)}$ via classical \\\tab optimization loop to find optimal ${\displaystyle \ket{\chi_i}}$.\\
    ${\displaystyle \hat{O} = f(\hat{A}, \hat{D}_{u}, \hat{D}_{\chi_1}, ..., \hat{D}_{\chi_{N}})}$};

\node at (10.7, -1.1) {\textbf{(c)} \underline{Typical Expectation-Value Circuit} };
\node at (12.2, -1.6) {$\braket{\psi|\hat{D}_{\phi}|\tilde{\psi}}$};

\node[scale=0.85] at (11.7,-3.7) {
\begin{quantikz}[column sep=0.15cm, row sep={0.5cm,between origins}]
    \lstick[wires=1]{$\ket{0}$} & \gate{H} & \ctrl{1} & \ctrl{6} & \ctrl{8} & \ctrl{7} & \ctrl{6} & \ctrl{5} & \ctrl{1} & \gate{H} & \meter{} \\
    \lstick[wires=4]{$\ket{0}^{\otimes n}$} & \qw & \gate[wires=4]{\hat{U}(\tilde{\boldsymbol{\lambda}})} & \qw & \qw & \qw & \qw & \ctrl{4} & \gate[wires=4]{\hat{U}^\dag({\boldsymbol{\lambda}})} & \qw & \qw \\
    \qw & \qw & \qw & \qw & \qw & \qw & \ctrl{4} & \qw & & \qw & \qw\\
    \qw & \qw & \qw & \qw & \qw & \ctrl{4} & \qw & \qw & & \qw & \qw\\
    \qw & \qw & \qw & \qw & \ctrl{4} & \qw & \qw & \qw & & \qw & \qw\\
    \lstick[wires=4]{$\ket{0}^{\otimes n}$} & \qw & \qw & \gate[wires=4]{\hat{U}(\boldsymbol{\theta})} & \qw & \qw & \qw & \targ{}  & \qw & \qw & \qw\\
    \qw & \qw & \qw & \qw & \qw & \qw & \targ{} & \qw & \qw & \qw & \qw\\
    \qw & \qw & \qw & \qw & \qw & \targ{} & \qw & \qw & \qw & \qw & \qw\\
    \qw & \qw & \qw & \qw & \targ{} & \qw & \qw & \qw & \qw & \qw & \qw
  \end{quantikz}
  };

\end{tikzpicture}
\caption{Algorithmic Framework. \textbf{(a)} Quantum variational PDE solving workflow. $\hat{U}$ is an ansatz gate, and $\lambda_0$ and $\boldsymbol{\lambda}$ are the variational parameters to be trained to represent our function of interest, with a tilde denoting the previous timestep's parameters. $C_u$ is the main cost-function to be minimized, returning the optimal state after one timestep. First, read in initial ansatz parameters to represent $u(t_i)$. Decompose the discretized (nonlinear) differential operator $\hat{O}$ into Adder and Diagonal gates. By training intermediate quantum states, a much broader class of nonlinearities can be represented than possible through naive applications of the Adder and Diagonal gates as considered in Ref.~\cite{lubasch}. Then use $\hat{O}$ to construct a cost-function which is then optimized to find $\ket{u}$ from $\ket{\tilde{u}}$, as in Ref.~\cite{lubasch}. Here, the cost-function shown represents the Backward Euler discretization scheme, though it can easily be adapted to other schemes. Repeat for desired number of timesteps, and then extract the desired classical data from the solution at the final timestep $\ket{u(t_f)}$. \textbf{(b)} Decomposition of the operator $\hat{O}$ into Adder and Diagonal gates via training intermediate quantum states. $\theta_0^i$ and $\boldsymbol{\theta^i}$ are the variational parameters corresponding to intermediate states $\ket{\chi^i}$, which are utilized in Diagonal gates $\hat{D}_{\chi_i}$to synthesize the full nonlinear operator $\hat{O}$. Cost-functions $C_{\chi_i}$ are iteratively constructed utilizing the Adder gate and the Diagonal gates of states already trained, which are then minimized to return the optimal intermediate parameters to represent any necessary nonlinearities / inhomogeneous terms. The number of intermediate states $N$ and the functions $f^k, f$ will depend on the exact form of $\hat{O}$.\textbf{(c)} Example of the estimation of a typical expectation value appearing in the cost-functions via the Hadamard test. $\ket{\psi}$ and $\ket{\phi}$ represent the normalized versions of $\ket{u}$ and $\ket{\chi}$, respectively.}
\label{fig:framework}
\end{figure*}
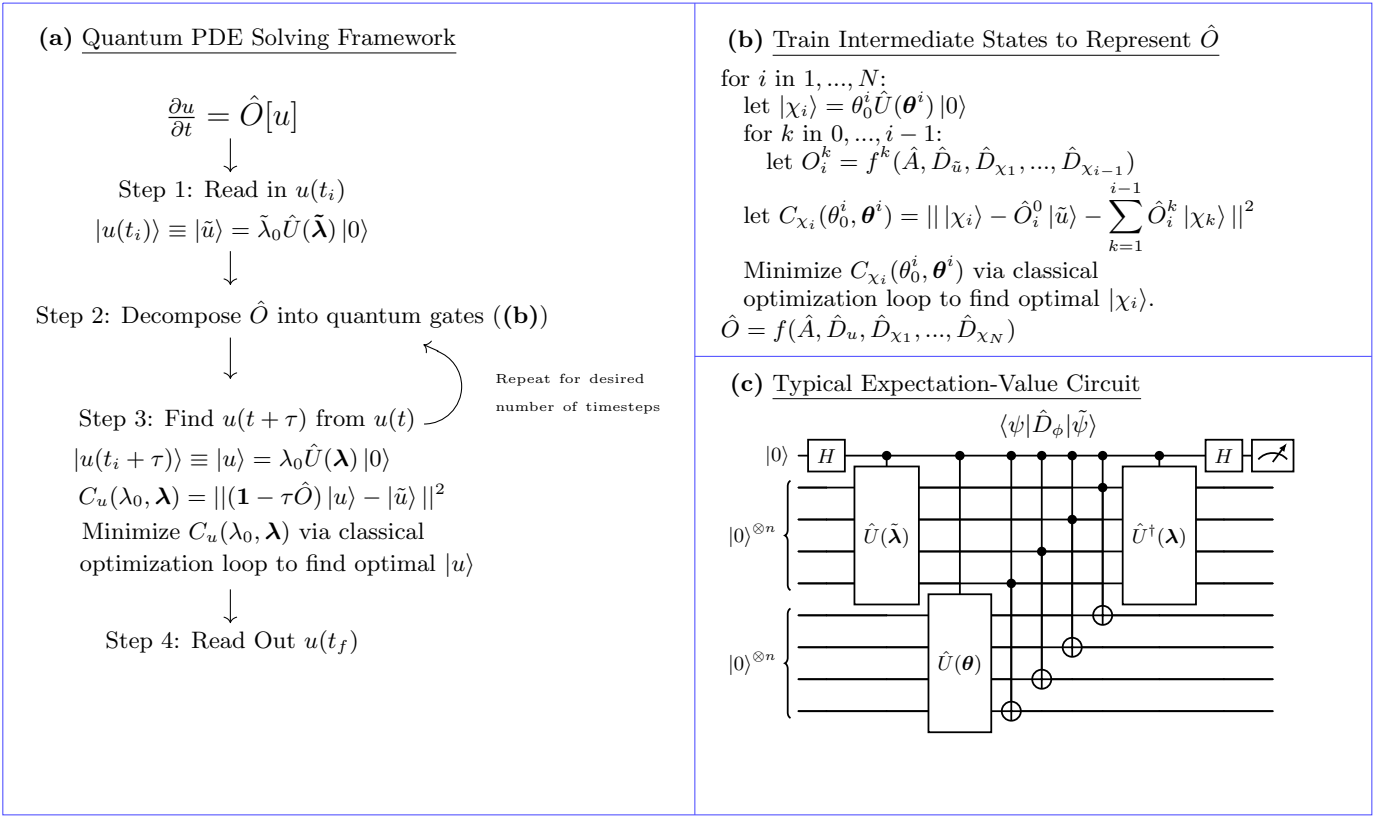


\section{General Algorithmic Framework}
\label{sec:method}

Consider a general PDE which is first order in time, $\frac{\partial u}{\partial t}=\hat{O}[u]$, where $\hat{O}$ is some potentially nonlinear differential operator depending on $u$, with an initial condition $u(x, t=0) = u_0(x)$. Our methodology, based on that of Lubasch et al. \cite{lubasch}, is to represent the solution $u(x, t)$ as a quantum state $\ket{u(t)} = \sum_{k=0}^{2^n-1}u(x_k,t)\ket{k}$, where $n$ is the number of qubits. In other words, $\ket{u(t)}$ has amplitudes in the computational basis consisting of the function $u(x, t)$ evaluated at each gridpoint $x_k$ in the discretized domain, of which there are a total of $N=2^n$. The problem of solving the PDE is analogous to finding the solution state after some small period of time $\tau$, $\ket{u(t+\tau)}$ using the solution state at time $t$ $\ket{u(t)}$. This state $\ket{u(t)}$ is represented by a low-depth parameterized gate $\hat{U}(\boldsymbol{\lambda})$, known as the ansatz, acting on the $\ket{0}$ state, and a cost-function is constructed which, when classically optimized, returns the optimal values of the parameters $\boldsymbol{\lambda} \in \mathbb{R}^{n_p}$ corresponding to the next timestep based on the values of the parameters at the current timestep. Since the function $u(t)$ is not necessarily normalized, we need an additional parameter $\lambda_0$, such that $\ket{u(t)}=\lambda_0\hat{U}(\boldsymbol{\lambda})\ket{0}=\lambda_0\ket{\psi}$, where we have defined $\ket{\psi}$ as the normalized state generated by the variational ansatz. To construct a cost-function, we first replace the time derivative with a finite difference approximation, e.g. the Backward Euler approximation, wherein our discretized PDE becomes $(\mathds{1}-\tau\hat{O})u(t+\tau)=u(t)$. Then, we define the cost-function as the square distance between both sides of this equation, $C_u = ||(\mathds{1}-\tau\hat{O})\ket{u(\lambda_0, \boldsymbol{\lambda})}-\ket{\tilde{u}(\tilde{\lambda}_0, \tilde{\boldsymbol{\lambda}})}||^2$, where we use a tilde to denote the previous timestep. The power of this method comes from the fact that this cost-function can be efficiently calculated on a quantum computer using the inner product, assuming that we can decompose $\hat{O}$ into quantum gates, as

\begin{equation} 
\begin{split}
    C(\lambda_0, \boldsymbol{\lambda}) = &\bra{\lambda_0(\mathds{1}-\tau\hat{O})\psi(\boldsymbol{\lambda})-\tilde{\lambda}_0\tilde{\psi}(\boldsymbol{\tilde{\lambda}})} \\  & {\lambda_0(\mathds{1}-\tau\hat{O})\psi(\boldsymbol{\lambda})-\tilde{\lambda}_0\tilde{\psi}(\boldsymbol{\tilde{\lambda}})} \rangle.
\end{split}
\label{costfunc}
\end{equation}

The problem remains of how to convert the differential operator $\hat{O}$ into a linear combination of quantum operators that can be implemented by standard quantum gates. Any spatial derivative term in $\hat{O}$ can be easily constructed after introducing the Adder operator, $\hat{A}$. As described in the supplemental material of Ref.~\cite{lubasch}, the Adder acts on a state $\ket{\psi}=\sum_{k=0}^{2^n-1}\psi_k\ket{k}$ by cyclically permuting it as such: $\hat{A}\ket{\psi}=\sum_{k=1}^{2^n}\psi_k\ket{k-1}$, where $\psi_{2^n}=\psi_0$. The conjugate $\hat{A}^\dag$ clearly has the opposite effect of cyclically permuting it the other direction. Equipped with this operator, we can approximate spatial derivatives using central finite differences, $\frac{\hat{\partial}}{\partial x} = \frac{2^{n-1}}{L}(\hat{A}-\hat{A}^\dag)$, $\frac{\hat{\partial^2}}{\partial x^2} = \frac{4^{n}}{L^2}(\hat{A}+\hat{A}^\dag-2\mathds{1})$, and so on for higher order derivatives, where $L$ is the length of the domain. Note that our use of the Adder operator imposes periodic boundary conditions on our solution, which may not always be desirable. However, various coordinate substitutions can be made to convert various other boundary conditions, such as Dirichlet conditions, into periodic boundary conditions. Further, Dirichlet boundary conditions may be approximated by periodic boundary conditions by doubling the spatial domain and reflecting the function about one of the original domain's endpoints, to create and time evolve a new function which is now initially periodic, as in Ref.~\cite{Gonzalez-Conde2021}, a method which we use to solve the Black--Scholes equation in sections V and VI. Recent papers have also introduced a method to recover Dirichlet conditions by calculating a small number of additional expectation values, as in Ref.~\cite{Sato2021}.

In order to deal with nonlinearities in $\hat{O}$, we must define the Diagonal operator, $\hat{D}$, introduced in Ref.~\cite{lubasch}. Given a state $\ket{\beta}=\sum_{k=0}^{2^n-1}\beta_k\ket{k}$, the Diagonal operator $\hat{D}_\beta$ acts on a state $\ket{\alpha}=\sum_{k=0}^{2^n-1}\alpha_k\ket{k}$ as such: $\hat{D}_\beta\ket{\alpha}=\sum_{k=0}^{2^n-1}\beta_k\alpha_k\ket{k}$. In other words, it acts as a point-wise multiplication operator between two quantum states; note that this is not a unitary transfomation on $\ket{\alpha}$, and as such, requires n dirty ancilla qubits to implement, where n is the number of qubits in $\ket{\beta}$ and $\ket{\alpha}$. Further, this means that $\hat{D}^\dag\hat{D}$ is not equal to identity. This operator can be constructed from any gate which generates $\ket{\beta}$, i.e. $\hat{U}_\beta$ such that $\hat{U}_\beta \ket{0} = \ket{\beta}$. This will allow us to generate nonlinear operators. Consider a general, potentially nonlinear operator of the form $\hat{O}[u] = f(x, u, \frac{\partial u}{\partial x}, ...)$, where $f$ is analytic in $u$ and it's derivatives. We need to construct a quantum state $\ket{f} = \hat{O}\ket{u}$, or, for an explicit method, $\ket{f} = \hat{O}\ket{\tilde{u}}$, which we can then inject into our cost-function to evaluate it from simple expectation values. As Lubasch et al. \cite{lubasch} point out, this is easy to do when $\ket{f}$ can be generated by repeated applications of Adders and Diagonal gates to $\ket{u}$, as is for operators which are polynomial in $u$ and linear in it's derivatives. For example, a nonlinear differential operator $\hat{O}[u] = u^2\frac{\partial u}{\partial x}$ has a quantum analogue $\ket{f} = \hat{O}\ket{u} = \hat{D}^2_{u}\frac{\hat{\partial}}{\partial x}\ket{u} = \frac{2^{n-1}}{L}\hat{D}^2_{u}(\hat{A}-\hat{A}^\dag)\ket{u}$. 

However, for more complicated nonlinearities, or operators with explicit $x$ dependence, it is not clear how one could construct $\ket{f}$ by directly applying operators to $\ket{u}$ or $\ket{\tilde{u}}$. The key is to circumvent this problem by representing $\ket{f}$ variationally in the same way as $\ket{u}$, by optimizing for parameters $\theta_0, \boldsymbol{\theta}$ to represent $\ket{f(\theta_0, \boldsymbol{\theta})}$ from the previous timestep's solution, $\ket{\tilde{u}}$. This can be done by breaking down $f(x, u, ...)$ into simple functions which can be represented by direct applications of Adder and Diagonal operators, and then synthesizing these simple functions back together into the full nonlinearity. Depending on how complicated the nonlinearity is, this may require several intermediate optimization steps. This makes the numerical scheme semi-implicit, whereby complicated nonlinearities are calculated explicitly from the previous timestep's solution before being injected into the implicit time evolution.

For example, consider a nonlinearity of the form $f(\frac{\partial u}{\partial x}) = \sin(\frac{\partial u}{\partial x})$. We expand $\sin$ into a truncated Taylor series, to find $f(\frac{\partial u}{\partial x}) \approx \frac{\partial u}{\partial x} - \frac{1}{6}(\frac{\partial u}{\partial x})^3 + \frac{1}{120}(\frac{\partial u}{\partial x})^5$. Let $\ket{\chi} = \theta_0\hat{U}(\boldsymbol{\theta})\ket{0} = \frac{\hat{\partial}}{\partial x} \ket{\tilde{u}}$. We can solve for $\theta_0, \boldsymbol{\theta}$ by optimizing the cost-function $C_\chi(\theta_0, \boldsymbol{\theta}) = ||\ket{\chi} - \frac{\hat{\partial}}{\partial x} \ket{\tilde{u}}||^2$. After doing so, we can fully construct $\ket{f(\theta_0, \boldsymbol{\theta})} = \ket{\chi} - \frac{1}{6}\hat{D}^2_\chi\ket{\chi} + \frac{1}{120}\hat{D}^4_\chi\ket{\chi}$, which can then be used to time evolve our PDE of interest. It is straightforward to construct operators with explicit $x$ dependence as well, by using quantum state preparation techniques to prepare the state $\sum_{k=0}^{2^n-1}f(x_k)\ket{k}$ corresponding to a generic function $f(x)$. A form of this technique for generating nonlinearities was used in Ref.~\cite{Mocz2021} for solving a nonlinear Poisson equation, and in Ref.~\cite{YewLeong2022} for Reaction-Diffusion equations. Here we have presented the most general version, whereby it is applicable to nearly all first order in time PDEs of interest in the literature. The method is straightforwardly adapted to other Runge-Kutta methods, such as the celebrated RK4 algorithm. In the following sections we demonstrate its accuracy for solving several differential equations with vastly different forms of nonlinearities.

It is obvious that this algorithm requires the ansatz parameters representing the initial condition $u(x, t=0)=u_0(x)$ a priori. We defer the discussion of how one can find the appropriate parameters to represent the initial condition to the next section. For optimization purposes we calculate all of the cost-functions appearing in this work using matrix multiplications to directly apply the effects of the Adder and Diagonal gates to avoid the computational overhead of having to explicitly simulate ancilla qubits, though we did simulate entire circuits with the Adder and Diagonal gates using Cirq for a few timesteps of each equation to ensure the results were the same as when using the computational shortcut. All of our expectation values in our simulations except for Section \ref{sec:hardware} were exactly calculated using this shortcut. 


\begin{figure*}[t]
     \centering
     \includegraphics[width=15cm]{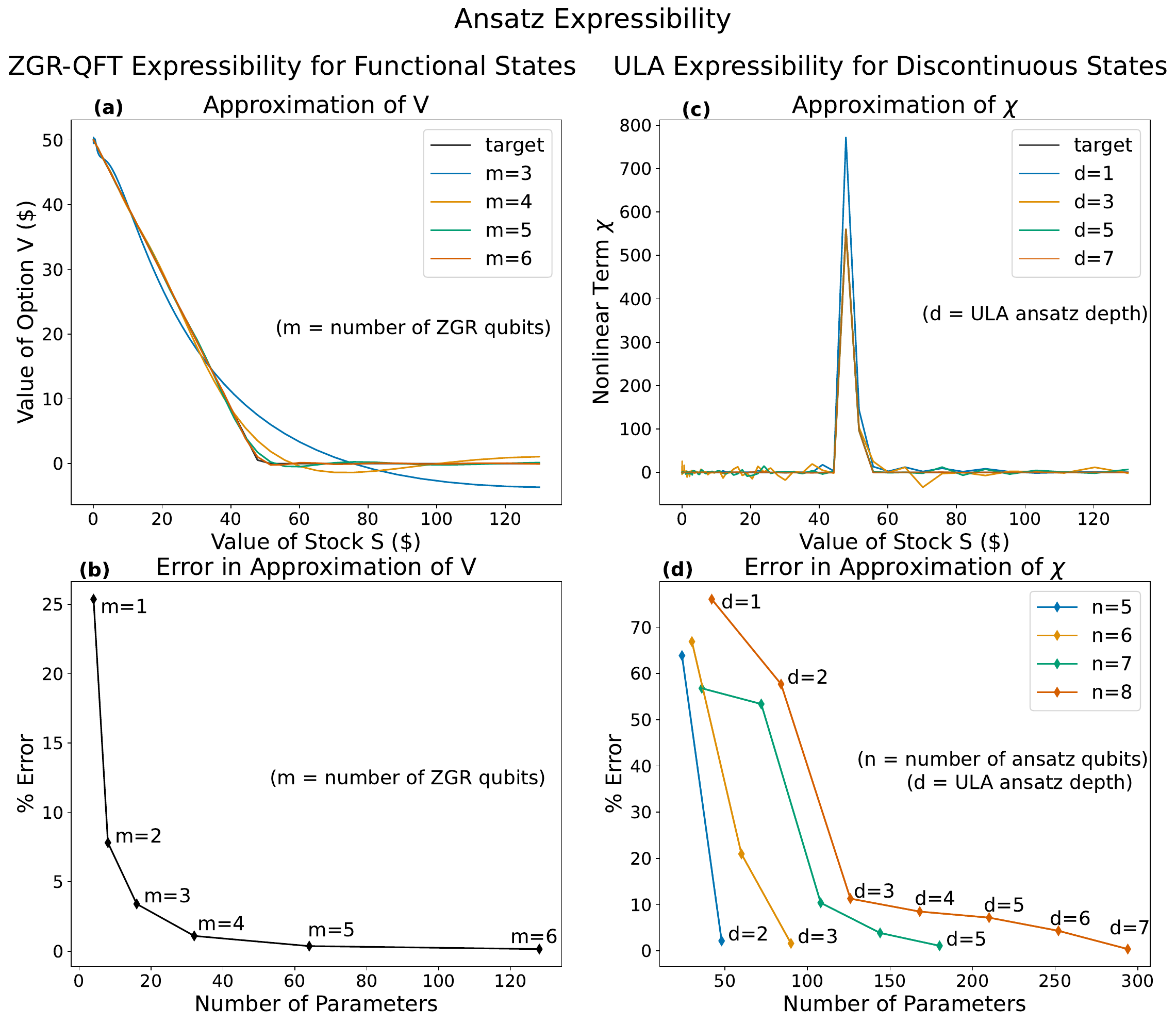}
     \hfill
        \caption{\textbf{Expressibility of Ansatz vs Number of Parameters}. Here, $n$ is the number of qubits, $m$ is the number of ZGR qubits for the ZGR-QFT, and $d$ is the depth for the ULA. $V$ represents the function of interest for the Black--Scholes problem (Eq.~\ref{nlbse}), while $\chi$ is an intermediate state used to construct the nonlinearity in the problem. The real-valued ZGR-QFT ansatz is able to accurately approximate the continuous terminal condition for $\ket{V}$ (the put option), and the Universal Layered Ansatz (ULA) is able to accurately approximate the sparse, discontinuous terminal condition for $\ket{\chi}$. We choose these terminal conditions as a representative example for the expressibility \cite{Cerezo2020-2} of the ansaetze because they are the least smooth functions considered in the paper. All simulations were conducted with $n=8$ ansatz qubits, except for the bottom right plot where $n$ varies. The ZGR-QFT has the useful property that its expressibility as a function of the number of parameters is invariant with respect to $n$, assuming that the function being approximated is continuous. This follows from the fact that, for a constant $m$, the ZGR-QFT will approximate the target distribution with the same number of Fourier coefficients regardless of $n$, meaning that the ansatz does not become more difficult to train as you increase the number of qubits. For the ULA, the necessary depth $d$ to get a good approximation of the target distrubution scales roughly linearly with the number of qubits, making the necessary number of parameters $O(n^2)$.}
    \label{fig:expressibility}
\end{figure*}


\section{Ansatz Selection}

One important element of this algorithm which has been ignored thus far is the form of the ansatz gate $U(\boldsymbol{\lambda})$. One needs an ansatz which is expressive enough to represent the solution states, the shape of which may not be known a priori, yet which is not too general as to impede its trainability. In particular, the ansatz needs to be able to represent functional states, but not general distributions. By functional states, we mean states whose amplitudes form a piecewise analytic function, as any solution to a PDE with piecewise analytic terms must be piecewise analytic. This greatly restricts the type of states we wish to be able to represent, and as such it allows us to create a more specialized ansatz which is well equipped to represent solutions to any PDE with piecewise analytic terms with a low number of parameters. In particular, we aim to represent solutions using a finite Fourier series, utilizing the powerful Quantum Fourier Transform (QFT). We do so using the so-called Real-Valued ZGR-QFT ansatz, shown in Appendix \ref{appendix:ZGRQFT}, an extension of the Fourier Series Loader introduced in Ref.~\cite{Watts2023} which parameterizes real-valued partial Fourier series. This ansatz allocates $m$ qubits to serve as representing Fourier coefficients, of which there are a total of $2^m$. These coefficients are fully parameterized using the ZGR parameterization, introduced in Ref.~\cite{Molina2022}. For a complex Fourier series, we could then apply the QFT to transform from momentum space to real space on all $n$ ansatz qubits, as done in \cite{Watts2023}. In order to get a real-valued Fourier series, we must instead apply various controlled versions of the ZGR gate, as well as an Adder and controlled Hadamards (see Appendix \ref{appendix:ZGRQFT}). Due to the ZGR, this ansatz has complexity $O(2^m)$; however, in most cases, we only need to choose $m \ll n$ to properly represent a functional state, as it rarely takes a very large amount of terms in a finite Fourier series to represent sufficiently smooth functions. Accounting for the other gates such as the QFT, the total complexity of the ansatz as $O(n\log(n)+2^m)$, with a number of rotation parameters equal to $2^{m+1}$. This ansatz has severable desirable properties. First, it is unchaotic, in the sense that smooth changes in parameters lead to smooth changes in the output distribution. This makes optimization much easier, as optimal parameters change relatively little between timesteps as compared to chaotic ansaetze (of which most popular ansaetze are, including the Alternating Layered Ansatz, Hardware Efficient Ansatz, etc.), and so they are easier to find for most classical optimizers. Second, there exist exact formulas for the parameters based on the target state you wish to represent, as described in Ref.~\cite{Mottonen2004}. These formulas can easily be extended to the full ZGR-QFT ansatz, as the QFT is equivalent to a classical Discrete Fourier Transform. This effectively makes the task of read-in, as in finding the ansatz parameters to represent the initial condition $f(x, t=0)$ (or, in the case of the Black--Scholes equation, the terminal condition $f(x, t=T)$), trivial. The inverse of these formulas also makes read-out, as in, the problem of extracting useful information from the solution state using the optimized ansatz parameters, trivial in that one can classically reconstruct the solution state directly from the ansatz parameters. In general, when using a different ansatz, for example, one can always make use of approximate polynomial-time function loading methods and the Swap test to train any variational ansatz to represent a target initial condition distribution for read-in. The problem of read-out is more complicated in the general case, as the quantities of interest one may want to read-out of the solution state depend on the problem at hand. However, recent work has been done on efficiently extracting moments of solutions \cite{Mocz2021} as well as using quantum machine learning for read-out \cite{LLoyd2022}. Lastly, for any choice of $m$, the ansatz has a constant number of parameters, regardless of $n$. This effectively means that for any choice of $m$, increasing $n$ does not make the ansatz more difficult to optimize, which is an exceedingly rare property for ansaetze. Note that for the real-valued ZGR-QFT, $m$ must be less than or equal to $n-2$. This makes the ansatz perform poorly for low numbers of qubits ($n<5$), as one does not have access to enough Fourier modes to adequately represent solution states. 

On the other hand, in some cases we do need to represent more jagged distributions which are not amenable to Fourier series. For the BSE in particular, the terminal condition is non-differentiable at a point and as such, its derivatives are discontinuous. Because of this, it is difficult to optimize the real-valued ZGR-QFT ansatz to represent $\ket{\chi}$ (defined in section V). Instead, we employ the Universal Layered Ansatz (ULA), shown in Appendix \ref{appendix:ULA}, which is a real-valued ansatz that is similar in form to the popular Alternating Layered Ansatz (ALA) \cite{ALA}, but which is optimally efficient in the sense that it can fully parameterize a real n-qubit state with $\frac{2^n(2^n-1)}{2}$ parameters, which is the minimum number of parameters needed to parameterize $SO(2^n)$. Further, it is hardware efficient, in that it only contains nearest-neighbor gates. This ansatz has a number of parameters equal to $6(n-1)d$, where $d$ is the number of layers of the ansatz. When the states of interest have low entanglement, we expect that we can take $d$ to be small. In practice, we find that the ansatz has enough expressibility to represent the states of interest ($\ket{\chi}$) when $d$ is of order $O(n)$, giving a total number of parameters and gate depth of $O(n^2)$. In addition to representing very jagged distributions, the ULA tends to outperform the real-valued ZGR-QFT ansatz for low ($n<5$) numbers of qubits. Figure \ref{fig:expressibility} shows how the expressibility of the ZGR-QFT and the ULA ansaetze scales with the number of parameters. It is suspected that ALA-like ansatze are difficult to simulate classically. However, the many desirable properties of the ZGR-QFT ansatz also make it classically tractable, in the sense that circuits involving only the ZGR-QFT ansatz can be classically simulated in polynomial time. Thus, the ZGR-QFT ansatz alone is not enough to build a nontrivial quantum PDE solving algorithm. It remains a crucial open problem to find an ansatz with similar desirable properties while also remaining classically intractable. 


\begin{figure*}[t]
     \centering
     \includegraphics[width=17cm]{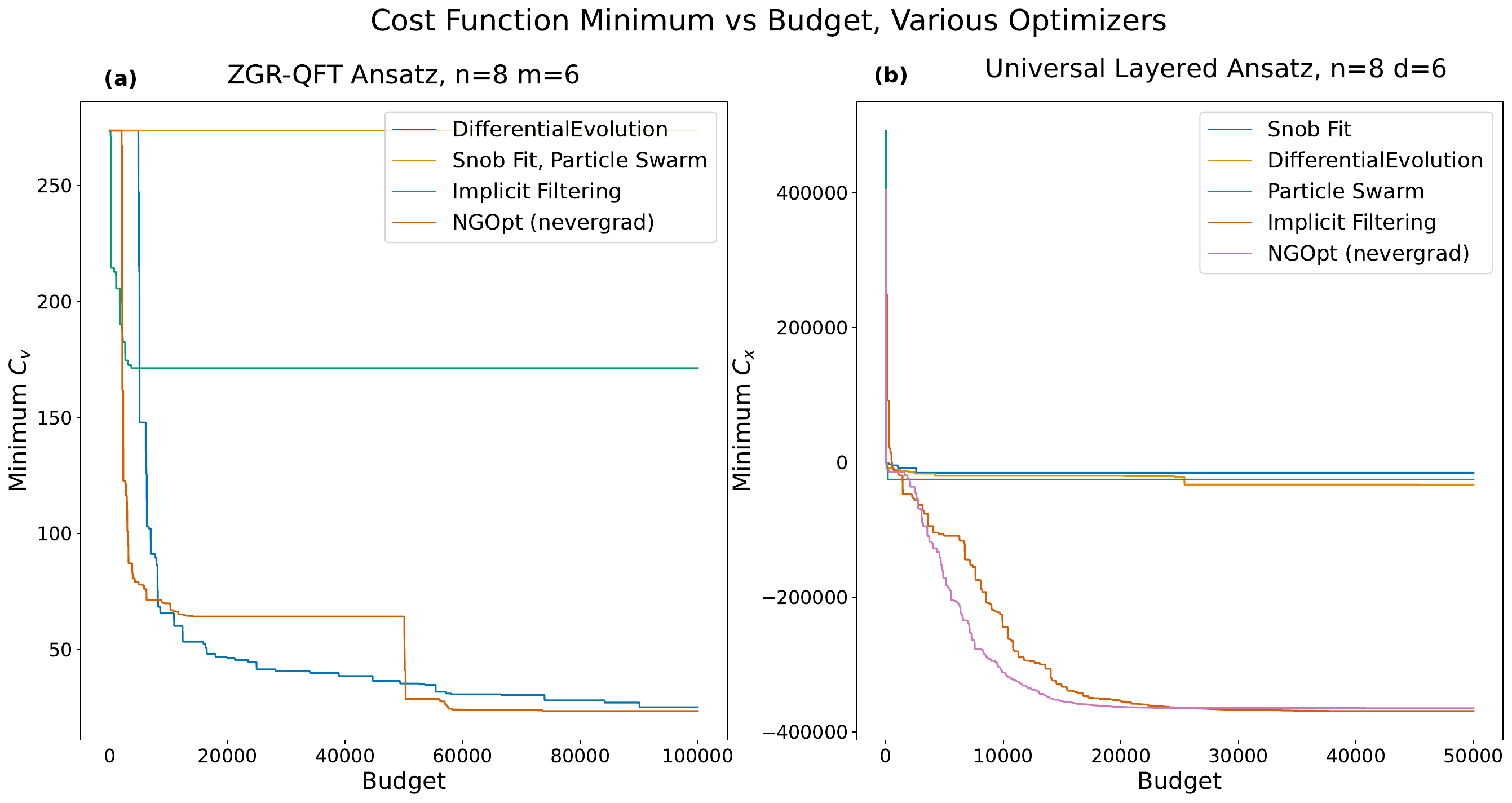}
     \hfill
        \caption{\textbf{Best Cost-Function Value Found by Optimizer vs Budget} Here \textit{budget} refers to the total number of cost-function evaluations that the optimizer software is allowed to make per PDE timestep in the PDE solving (Step~3 in Fig.~\ref{fig:framework}). (In our circuit simulations, to evaluate the cost function, we directly computed expectation values, so this plot has nothing to do with the number of times a circuit would need to be executed on quantum hardware to estimate each expectation values.) This figure shows how difficult it is for different optimizers to optimize the (a) ZGR-QFT ansatz (b) and ULA. Both anszaete were set to use $n=8$ qubits, with $m=6$ for the ZGR-QFT and $d=6$ for the ULA. For the ZGR-QFT only Differential Evolution and NGOpt converged to the global minimum, and for the ULA only NGOpt and Implicit Filtering reached the global minimum. The optimizations were performed for the first timestep in solving the nonlinear Black--Scholes equation.}
     \label{fig:optimizers}
\end{figure*}



\section{Optimization}

To optimize the cost-functions appearing in this work, we use the previous timestep's parameters as the initial guess for the new timestep's parameters, and we conduct a line search on the norm parameter $\lambda_0$ with a very low budget before running a standard optimization algorithm on all of the parameters. We tested several different popular gradient-free optimizers (Differential Evolution, Implicit Filtering, Particle Swarm, Snob Fit, and an adaptive optimizer offered by the nevergrad python package called NGOpt) on the first timestep of the nonlinear Black--Scholes equation (Eq.~\ref{nlbse}) for both cost-functions and ansaetze, shown in Figure~\ref{fig:optimizers}. The previous timestep's parameters were used as the initial guess of the new parameters for each optimizer. We choose the first timestep as a representative case to test the optimization for the whole algorithm because the nonlinearity is largest at the beginning of the time evolution. We also tested some popular gradient-based optimizers (BFGS, Truncated Newton's Method, and ADAM), though all of them required a budget (maximum number of cost-function evaluations per timestep) at least on the order of $p \cdot 10^4$ to perform well, where $p$ is the number of ansatz parameters, which we found to be too high to extract useful results. This is not particularly surprising as we expect that our cost functions may suffer from the so-called barren plateau problem \cite{Cerezo2020-1}. Further work may be done to investigate and mitigate the existence of barren plateaus in our cost functions. For our full simulation of the algorithm we use Differential Evolution for optimizing the ZGR-QFT ansatz and NGOpt for optimizing the Universal Layered Ansatz. Because the ZGR-QFT ansatz is smooth, we can place relatively tight bounds on the rotation parameters while optimizing them, as most of them only change on the order of $10^{-1}$ or $10^{-2}$.

\section{Solving The Barles and Soner Nonlinear Black--Scholes Equation}
\label{sec:bse}

\subsection{Cost-functions, Circuits, Complexity}

The Black--Scholes equation is a practically important PDE to solve in finance -- it models the price of an option as a function of time and the underlying asset price \cite{QuantFinance}. Denoting $V$ as the option price and $S$ as the underlying asset price, the Black--Scholes model with the Put-option terminal condition can be simplified by making the substitution $x=\log(S)$ \cite{Gonzalez-Conde2021}, reducing it to the form:

\begin{equation}
\begin{gathered}
    \frac{\partial V}{\partial t} = rV+(\frac{\sigma^2}{2}-r)\frac{\partial V}{\partial x}-\frac{\sigma^2}{2}\frac{\partial^2 V}{\partial x^2} \\
    V(x, t=T) = \max(K-e^{x}, 0)\\
    V(x\rightarrow \infty, t) = 0.
\end{gathered}
\label{nlbse}
\end{equation}

\noindent Here, $r$ is the interest rate. The factor $\sigma$ is a model-dependent value known as the volatility. In the linear BSE, the volatility is taken to be a constant, leading to a Heat-like equation which can easily be analytically solved. However, more complicated and robust volatility models can have the volatility as a function of the option price, it's derivatives, the stock price, and time. These models necessarily lead to a nonlinear PDE which is often impossible to solve analytically. Of these models, one of particular interest is the Barles and Soner's model as described in Ref.~\cite{Ankudinova2008}, in which the volatility (squared) takes the form

\begin{equation}
    \sigma^2 = \sigma_0^2 \left[1+e^{r(T-t)}a^2 \Big( \frac{\partial^2 V}{\partial x^2}-\frac{\partial V}{\partial x} \Big)\right].
\end{equation}

\noindent From Eq.~\ref{nlbse} it is clear that $\hat{O} = r + (\frac{\sigma^2}{2}-r)\frac{\partial}{\partial x}+\frac{\sigma^2}{2}\frac{\partial^2}{\partial x^2}$ in this case. To convert this into a quantum operator, we must first generate an intermediate quantum state to represent $\chi = \frac{\partial^2 V}{\partial x^2}-\frac{\partial V}{\partial x}$, $\ket{\chi}$. After we have the state $\ket{\chi}$, the quantum operator takes the form 

\begin{align}
\begin{split}
    \hat{O} = r\mathds{1} + (\frac{\sigma_0^2}{2}(\mathds{1}+e^{r(T-t)}a^2\hat{D}_\chi)-r)\frac{\hat{\partial}}{\partial x}
    \\+\frac{\sigma_0^2}{2}(\mathds{1}+e^{r(T-t)}a^2\hat{D}_\chi)\frac{\hat{\partial}^2}{\partial x^2}.
\end{split}
\end{align}

This necessitates a two cost-function scheme, in which at each time step we first find the optimal variational parameters to represent $\ket{\chi}$ from the previous timestep's parameters for $\ket{\tilde{V}}$, and then we use those parameters to generate the Diagonal gate $\hat{D}_\chi$, which in turn we use to calculate the cost-function to find the optimal parameters to represent $\ket{V}$. We therefore also need a cost-function for $\ket{\chi}$, which can straightforwardly be calculated using the inner product distance method as before. Using the Backward Euler finite difference, our scheme becomes semi-implicit, with the cost-functions given by

\begin{equation}
\begin{gathered}
    C_\chi = ||\ket{\chi} - (\frac{\hat{\partial}^2}{\partial x^2}-\frac{\hat{\partial}}{\partial x})\ket{\tilde{V}}||^2, \\
    C_V = ||(\mathds{1}-\tau\hat{O})\ket{V}-\ket{\tilde{V}}||^2,
\end{gathered}
\end{equation}

\noindent where we have dropped the explicit parameter dependence of the cost-functions. One could equivalently define and minimize a joint cost-function $C_{\text{total}} \equiv C_{V}+\mu C_{\chi}$ for $\mu > 0$, though this makes optimization more difficult in practice. The full cost-functions can be found in Appendix \ref{appendix:1DBSE}. The equivalent cost-function for the 1D linear model in which the volatility is a constant is included in the appendix as well. The nonlinear cost-function for $V$ contains 19 expectation values which can be calculated via the Hadamard test. The Diagonal gate requires $n$ ancilla to generate, and the Adder can be diagonalized by the Quantum Fourier Transform (QFT) to be generated without any ancilla. The largest circuits are therefore the ones which contain two Diagonal gates. These circuits require $n$ ancilla for each Diagonal gate. In addition to the $n$ ansatz qubits and the control qubit, these circuits therefore have a total of $3n+1$ qubits. The Diagonal gate contains $O(n)$ gates, and the Adder contains $O(n\log(n))$ gates, due to the QFT. Therefore, the largest circuits contain $O(n)$ qubits and $O(n\log(n))$ gates, not accounting for the complexity of the ansatz(e). In general this will be true when applying the method to any equation. Our ansaetze as described in the previous section have at most $O(n^2)$ gates and $O(n^2)$ parameters, giving a total circuit complexity of $O(n^2)$ gates. As in Eq.~\ref{nlbse}, the BSE demands a Dirichlet condition on the right boundary, while the Adder imposes periodic boundary conditions. However, as aforementioned, the Dirichlet boundary condition can be approximated by a periodic boundary condition by doubling the length of the domain, reflecting the terminal condition across the right boundary, and time evolving the whole reflected function. This means that for $n$ ansatz qubits you get $2^{n-1}$ gridpoints representing your domain of interest, as one qubit must be used for the reflection. Note that this method does not perfectly capture the Dirichlet boundary conditions, and is a source of error accruing after several timesteps. Recently, methods have been proposed in the literature to impose Dirichlet boundary conditions exactly by calculating a small number of additional expectation values, as in Ref.~\cite{Sato2021}, though we were not aware of this method when running our simulations.


\begin{figure*}[t]
     \centering
        \includegraphics[width=17cm]{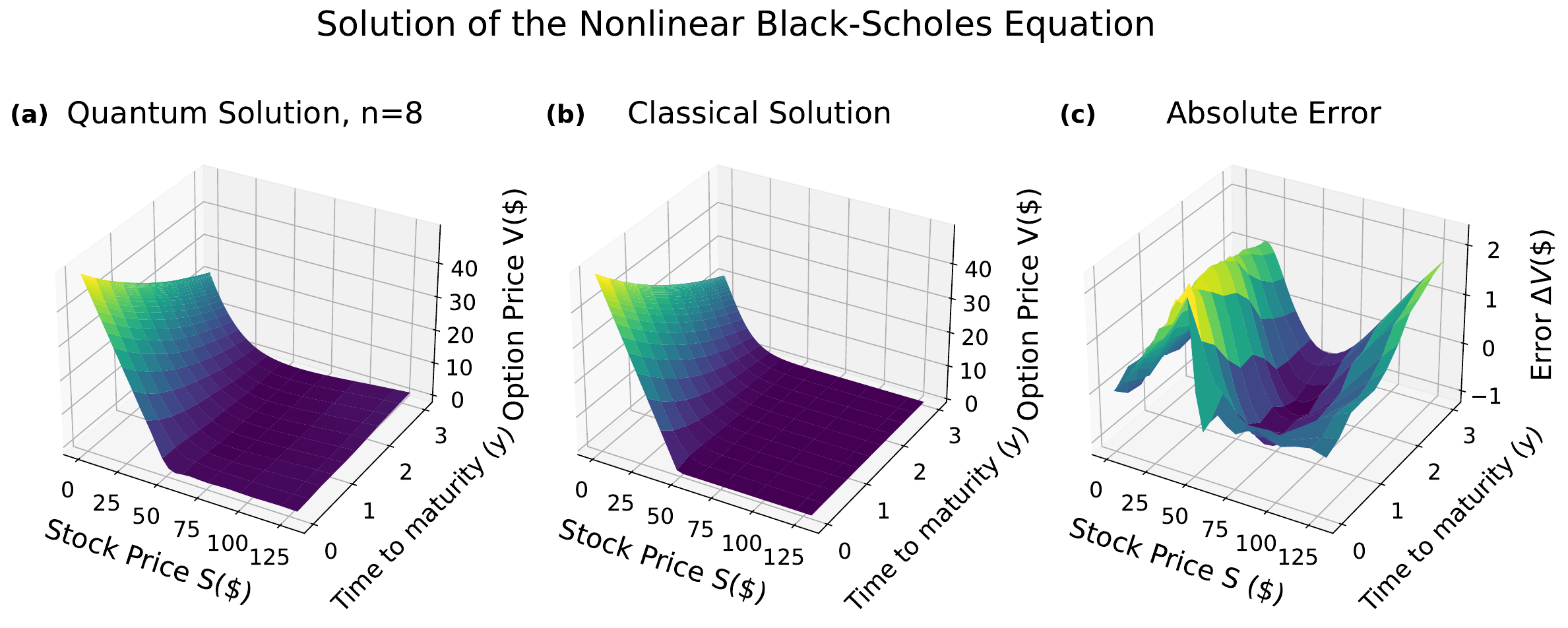}
        \caption{\textbf{1D Nonlinear Black--Scholes Equation Time Evolution}. The quantum and classical solutions closely align, with a final relative error of 2.36\% after the last timestep, calculated as $(V_\textrm{quantum} - V_\textrm{classical})/||V_\textrm{classical}||$, giving an error per timestep of 0.236\%. At the boundaries, most of the error accumulates due to the Gibbs phenomenon and the imperfect approximation of the Dirichlet boundary condition. There is also relatively large error due to the Gibbs phenomenon near $S=50$, where the terminal condition has a discontinuous derivative.}
    \label{fig:BSE}
\end{figure*}

\begin{figure*}[t]
    \centering
    \includegraphics[width=9cm]{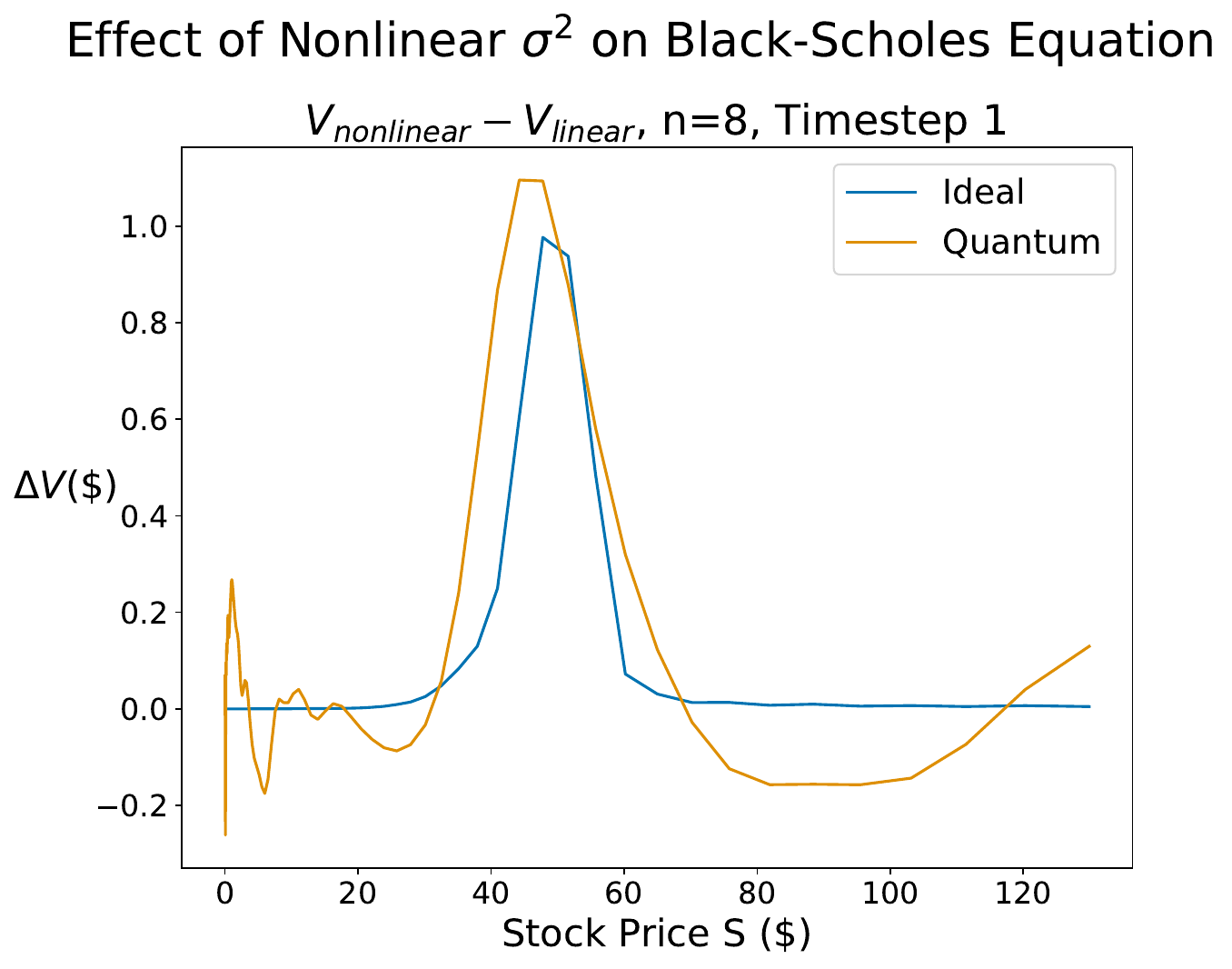}
    \caption{\textbf{Effect of Nonlinear Volatility ($\sigma^2$) in the Black--Scholes Equation (First Timestep)} The key feature of the nonlinear volatility (as described in Ref.~\cite{Ankudinova2008}) is that it causes the price of the option to sharply increase as compared to the linear model, peaked asymmetrically near the discontinuity in the terminal condition (around $S=50$). As evident from the plot, this qualitative behavior is reflected in the quantum solution, which differs from the ideal solution mostly due to edge effects, an artifact of the Fourier-based nature of the ansatz and the Gibbs phenomenon. The rest of the error stems from imperfect optimization. The ideal plot was calculated by classically numerically solving for the first timestep of the nonlinear Black--Scholes time evolution where the terminal condition is given by the (imperfect) ZGR-QFT representation of the Put option, and we use the reflected periodic boundary instead of Dirichlet boundary conditions so that errors associated with the optimization are not confounded with errors associated with the ``read-in'' of the terminal condition or the imperfect boundary condition.}
    \label{fig:BSEnonlinearity}
\end{figure*}



\subsection{Results}

For our simulations, we consider a European put-type option boundary value, where the option price at the maturity time $T$, which we choose to be $T=3$ years, is given as $V(S, T) = \max(K-S, 0)$. As in Ref.~\cite{Gonzalez-Conde2021}, we choose a domain of $S\in[\frac{1}{135}, 135]$, giving a transformed domain of $x\in[-\log(135), \log(135)]$, with parameter values as such: $K=50$, $r=0.3$, and $\sigma_0^2=0.04$. We also set the nonlinear parameter $a=0.1$. We begin with the transformed European call option $V(x, T) = \max(K-e^x, 0)$ and time evolve backwards from $t=T$ to $t=0$, with 10 timesteps of $\tau=-0.3$. We simulate 8 noiseless ansatz qubits, though 1 of them is used to reflect our function to deal with the Dirichlet boundary conditions (see Ref.~\cite{Gonzalez-Conde2021}), giving us a reflected transformed domain of length $L=4\log{135}$. Our target function is represented by a grid mesh of $2^{(8-1)}=128$ points. We use the ZGR-QFT ansatz to represent $\ket{V}$ with $m=6$, giving 129 total parameters, and the Universal Layered Ansatz to represent $\ket{\chi}$ with $d=6$, giving 253 total parameters. For comparison we also solved the linear and nonlinear BSE with the same parameters using a classical numerical solver on 128 grid points, which we constructed using the same semi-implicit numerical scheme used for our quantum solver and a classical matrix inversion algorithm. In Figure \ref{fig:BSE}, we employ the classical solver using the exact terminal condition for $V$ and the exact Dirichlet boundary conditions to get the ideal results for ``true" time evolution. In Figure \ref{fig:BSEnonlinearity}, we employ the classical solver using the imperfect Fourier approximation of the terminal condition and the imperfect (periodic) boundary condition to isolate the errors which accrue as a result of time evolution. As can be seen from the figures, the behavior of the time evolution is the same in the classical and the quantum simulations, showing that our method effectively captures the nonlinear dynamics of the equation in the noiseless setting. We find a total error of 2.36\% at the end of the time evolution as compared with the classical solution, which mostly stems from the approximate boundary condition; there is around 0.13\% error which arises before the time evolution due to the imperfect ansatz representation of the terminal condition.

\section{Solving the 2D Linear Black--Scholes Equation}

\subsection{2D Extension}
Consider the 2D linear Black--Scholes equation,
\begin{equation}
\begin{gathered}
    \frac{\partial V}{\partial t} = rV+(\frac{1}{2}\sigma_x^{2}-r)\frac{\partial V}{\partial x}+(\frac{1}{2}\sigma_y^{2}-r)\frac{\partial V}{\partial y} \\-\frac{1}{2}\sigma_x^2\frac{\partial^2 V}{\partial x^2}-\frac{1}{2}\sigma_y^2\frac{\partial^2 V}{\partial y^2}--\frac{1}{2}\rho\sigma_x\sigma_y\frac{\partial^2 V}{\partial x \partial y},\\
    V(x, y, t=T) = \max(K-w_x e^{x}-w_y e^{y}) \\
    V(x \rightarrow \infty, y, t) = 0 \\
    V(x, y \rightarrow \infty, t) = 0
\end{gathered}
\end{equation}

\noindent where $\sigma_x$ and $\sigma_y$, are the (constant) volatilities of the asset prices $S_1$ and $S_2$ respectively, $\rho$ is the cross-correlation of $S_1$ and $S_2$, and $w_x$ and $w_y$ are constant positive weights. Above we have made the substitutions $x=\log(S_1), y=\log(S_2)$. Since this equation is linear, we choose the (fully implicit) Backward Euler method for its numerical stability.

In order to generalize the method of Section II to two dimensions, we must first decide how to encode our solution state as a quantum state. We wish to partition our 2D domain into a grid of $2^{n_x}\times 2^{n_y}$ gridpoints, where we designate the first $n_x$ qubits to the $x$ dimension, and the last $n_y$ qubits to the $y$ dimension, with a total of $n=n_x+n_y$ ansatz qubits. In this form, the solution state takes the form $\ket{u(t)} = \sum_{k=0}^{2^n-1}u(x_{k_x}, y_{k_y}, t)\ket{k}$, where $k_y=k \mod 2^{n_y}$ spans from $[0, 2^{n_y}-1]$ and $k_x = \lfloor \frac{k}{2^{n_y}} \rfloor$ spans from $[0, 2^{n_x}-1]$, as expected. In other words, the solution state has amplitudes corresponding to the flattened 2D array of $u(x, y, t)$ evaluated at each point in the discretized domain, where the $x$-index is treated as the more significant index (one could just as well choose the $y$-index to be the more significant index, though we choose the former convention here). As before, we wish to represent $\ket{u}$ variationally, so that $\ket{u(t)}=\lambda_0\hat{U}(\boldsymbol{\lambda})\ket{0}=\lambda_0\ket{\psi}$. The 2D Diagonal operator (which we do not need for the 2D linear Black--Scholes equation) is the same as the 1D Diagonal operator. To convert the partial derivatives $\frac{\partial}{\partial x}$ and $\frac{\partial}{\partial y}$ and any higher-order spatial derivatives into quantum operators, we need only introduce the so-called $x$-Adder and $y$-Adders, 
$$
\hat{A}_x = \hat{A}_{n_x}\otimes\mathds{1}^{\otimes n_y}, \hat{A}_y = \mathds{1}^{\otimes n_x}\otimes \hat{A}_{n_y}
$$

\noindent where $\hat{A}_{n_x}$ denotes the previously discussed Adder operator acting on $n_x$ qubits. In other words, the $x$ and $y$ Adders are just the Adder operator acting on the $x$ qubits and the $y$ qubits, respectively. From these, the partial derivatives can be decomposed in terms of these Adders completely analogously to the 1D case, and the cost-function can be derived in the same way as well, which is shown in Appendix \ref{appendix:2DBSE}. The circuit complexities are the same as in the 1D case. This flattened array method of encoding 2D functions and constructing derivatives from tensor products of Adders with the identity generalizes straightforwardly to arbitrary dimensions, and even vector or tensor fields, allowing this quantum PDE solving method to be applied to even the Navier--Stokes Equations, demonstrating the immense generality of this method. For solving the 2D BSE, we choose our ansatz to be the straightforward generalization of the (complex) ZGR-QFT ansatz to 2 dimensions (see Appendix C of \cite{Watts2023}), which utilizes a finite 2D Fourier series in the same way that the 1D version utilizes a finite Fourier series. Here, we do not need to restrict ourselves to a real-valued Fourier series because our PDE of interest is linear, and thus we can just take the real part of our solution at the end. Using $n_x=n_y=6$ qubits for the $x$ and $y$ dimensions and $m_x=m_y=3$ qubits for the Fourier coefficients in each dimension, the ansatz has a total of $2^{m_x+m_y+3} = 512$ parameters.

\subsection{Results}
\begin{figure*}[t]
     \centering
     \includegraphics[width=17cm]{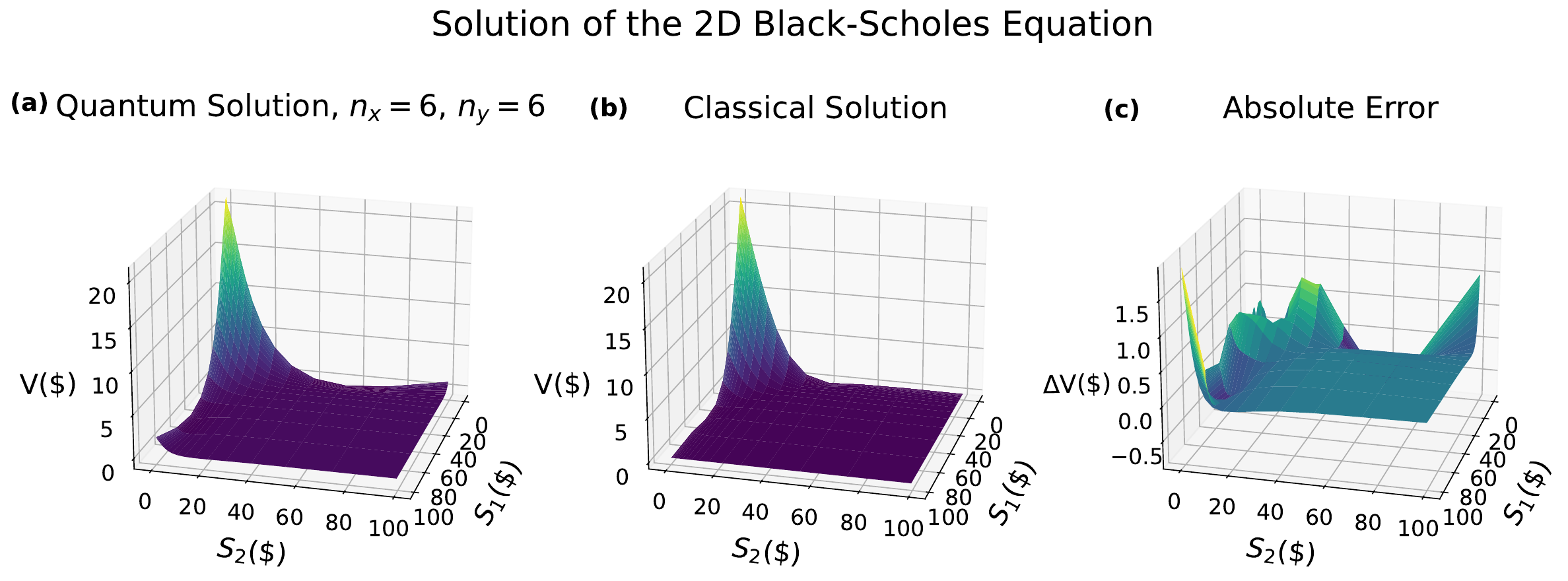}
        \caption{\textbf{2D Linear Black--Scholes Equation Solution (Final Timestep)}. The classical solution is calculated using the Backward Euler scheme with the exact initial condition. The quantum solution has an error of 2.60\% compared with the classical, giving an error per timestep of 0.260\%. As in the 1D case, most of the error is near the boundaries due to the approximate boundary conditions.}
    \label{fig:2DBSE}
\end{figure*}

For our numerical simulations, we let $r=0.3, \sigma_x=\sigma_y=0.2, w_1=w_2=1, \rho=0$, and once again time evolve backwards from $t=3$ to $t=0$ in 10 timesteps of $-0.3$. We set $n_x=n_y=6$, but one of each is used to reflect the function to deal with boundary conditions as in the 1D case, giving us a discretized domain of $32\times32=1024$ gridpoints. We found a final error of 2.60\%, and the results are depicted in Figure \ref{fig:2DBSE}. Like the 1D case, there is a small amount of error associated with read-in, and most of the error comes from the approximation of the boundary conditions.


\section{Solving the Buckmaster Equation}

\begin{figure*}[t]
    \centering
     \includegraphics[width=17cm]{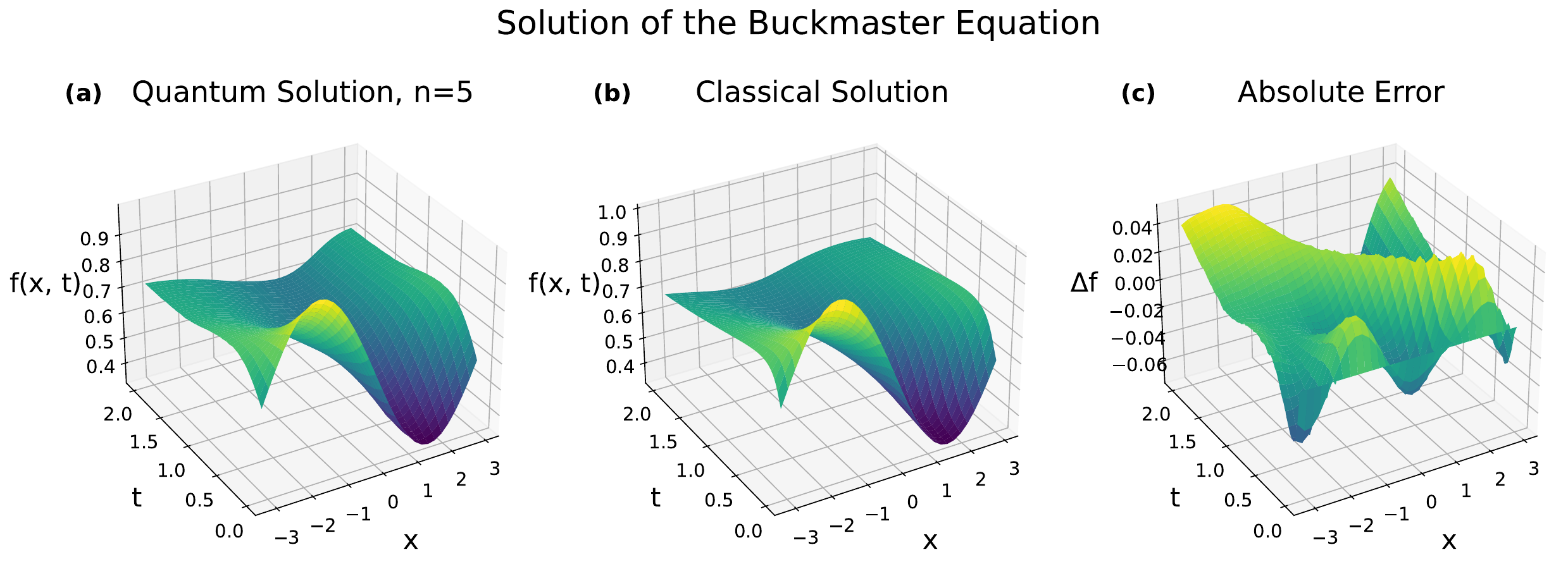}
     \caption{\textbf{Buckmaster Equation Time Evolution}. The classical solution is calculated using the Forward Euler scheme with the exact initial condition. The quantum solution has an error of 6.54\% compared with the classical, giving an error per timestep of 0.0262\%. The error is large near the boundaries despite the periodic boundary condition because any errors which accrue naturally near the boundaries as a result of the imperfect initial condition / optimization will grow larger over time due to the Gibbs phenomenon.}
     \label{fig:buckmaster}
\end{figure*}

Consider the Buckmaster equation,

\begin{equation}
     \frac{\partial f}{\partial t} = \frac{\partial^2}{\partial x^2}(f^4) + \alpha \frac{\partial}{\partial x}(f^3),
\end{equation}

\noindent where $\alpha$ is a known parameter. This equation governs the time evolution of the surface of a viscous liquid, a problem of relevance to fluid dynamics and for which it is difficult to find analytical solutions. Unlike the nonlinear Black--Scholes equation, the time evolution of the Buckmaster equation is entirely nonlinear, in that it cannot be approximated as a perturbation of a linear equation. For this reason we choose to verify the quantum PDE method on this problem. For simplicity we consider a fully explicit Forward Euler scheme, which requires less expectation values, allowing us to more quickly time evolve our solution state and verify the robustness of the algorithm in the many timestep regime. The procedure for deriving the cost-function for the Buckmaster equation is the same as described in Section II and the example given in Section V.A, so we omit discussion of the cost-function, though it can be found in Appendix \ref{appendix:Buckmaster}. For our simulations we let $\alpha=1$, and consider a domain of $x\in[-\pi, \pi]$ ($L=2\pi$) with a sinusoidal initial condition $f(x, t=0) = \frac{1}{3}(2-\sin(x))$ and periodic boundary conditions, so that we need not use the reflection trick to deal with the boundary conditions. We time evolve the equation for $n=5$ qubits and therefore a mesh of $2^5=32$ gridpoints from $t=0$ to $t=2$ in 250 timesteps of size 0.008, using the real ZGR-QFT ansatz with $m=3$. The results are shown in Figure \ref{fig:buckmaster}, in which the quantum solution completely captures the nonlinear time dynamics of the equation.


\section{Solving the Deterministic KPZ Equation}


\begin{figure*}[t]
     \centering
     \includegraphics[width=17cm]{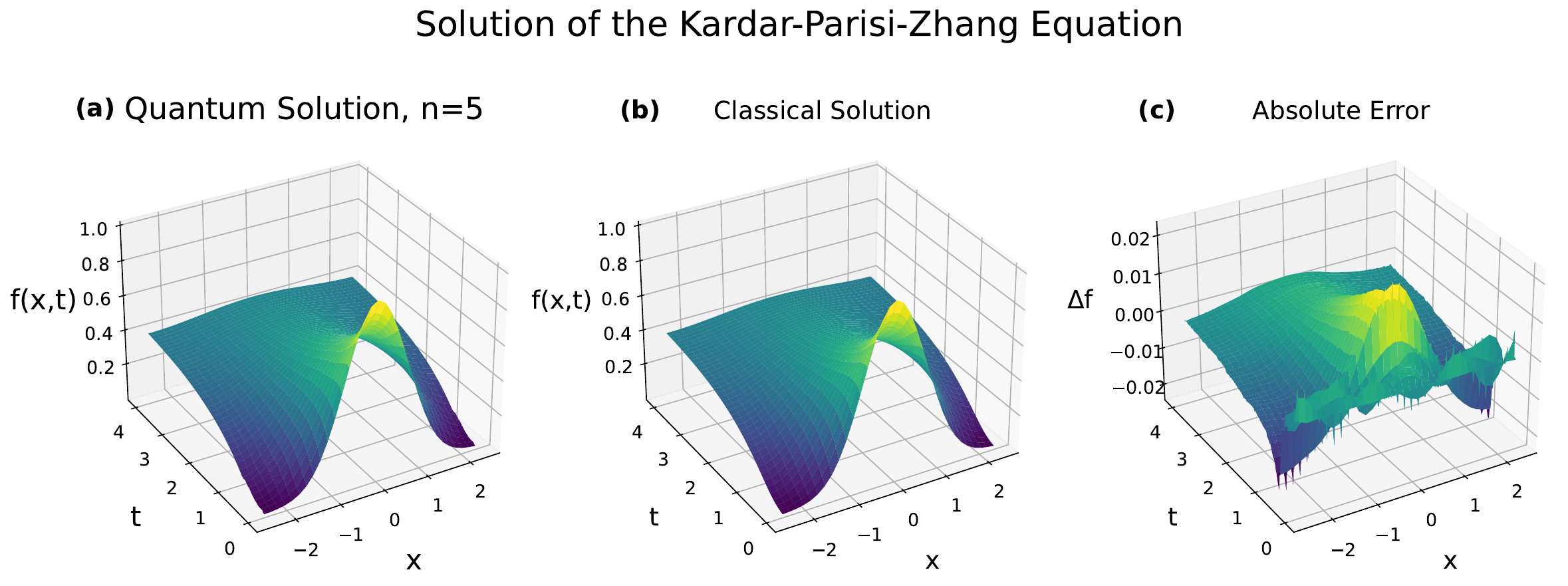}
     \caption{\textbf{KPZ Equation Time Evolution}. The classical solution is calculated using the Forward Euler scheme with the exact initial condition. The quantum solution has an error of 0.415\% compared with the classical, giving an error per timestep of 0.00208\%.}
     \label{fig:KPZ}
\end{figure*}


Lastly, consider the Deterministic Kardar--Parisi--Zhang (KPZ) equation, used to model deterministic surface growth:

\begin{equation}
     \frac{\partial f}{\partial t} = \alpha\frac{\partial^2 f}{\partial x^2} + \beta (\frac{\partial f}{\partial x})^2,
\end{equation}

\noindent where $\alpha$ and $\beta$ are given. The Deterministic KPZ equation describes the asymptotic behavior of a large class of deterministic surface growth models \cite{Chatterjee2021}. We again use a fully explicit Forward Euler scheme. The nonlinear term in the KPZ equation requires one auxiliary quantum state to be trained, as in the case of the nonlinear Black--Scholes equation, and a full description of the cost-functions can be found in Appendix \ref{appendix:KPZ}. For our simulations we let $\alpha=0.5$ and $\beta=0.5$, and consider a domain of $x\in[-2.5, 2.5]$ ($L=5$) with a Gaussian initial condition $f(x, t=0) = e^{-x^2}$ and periodic boundary conditions, so that we need not use the reflection trick to deal with the boundary conditions. We time evolve the equation for $n=5$ qubits and therefore a mesh of $2^5=32$ gridpoints from $t=0$ to $t=4$ in 200 timesteps of size 0.02, using the ZGR-QFT ansatz with $m=3$. The results are shown in Figure \ref{fig:KPZ}, and once again the quantum algorithm performs well.

\section{Execution on Hardware}
\label{sec:hardware}

\begin{figure*}[t]
    \centering
     \includegraphics[width=17cm]{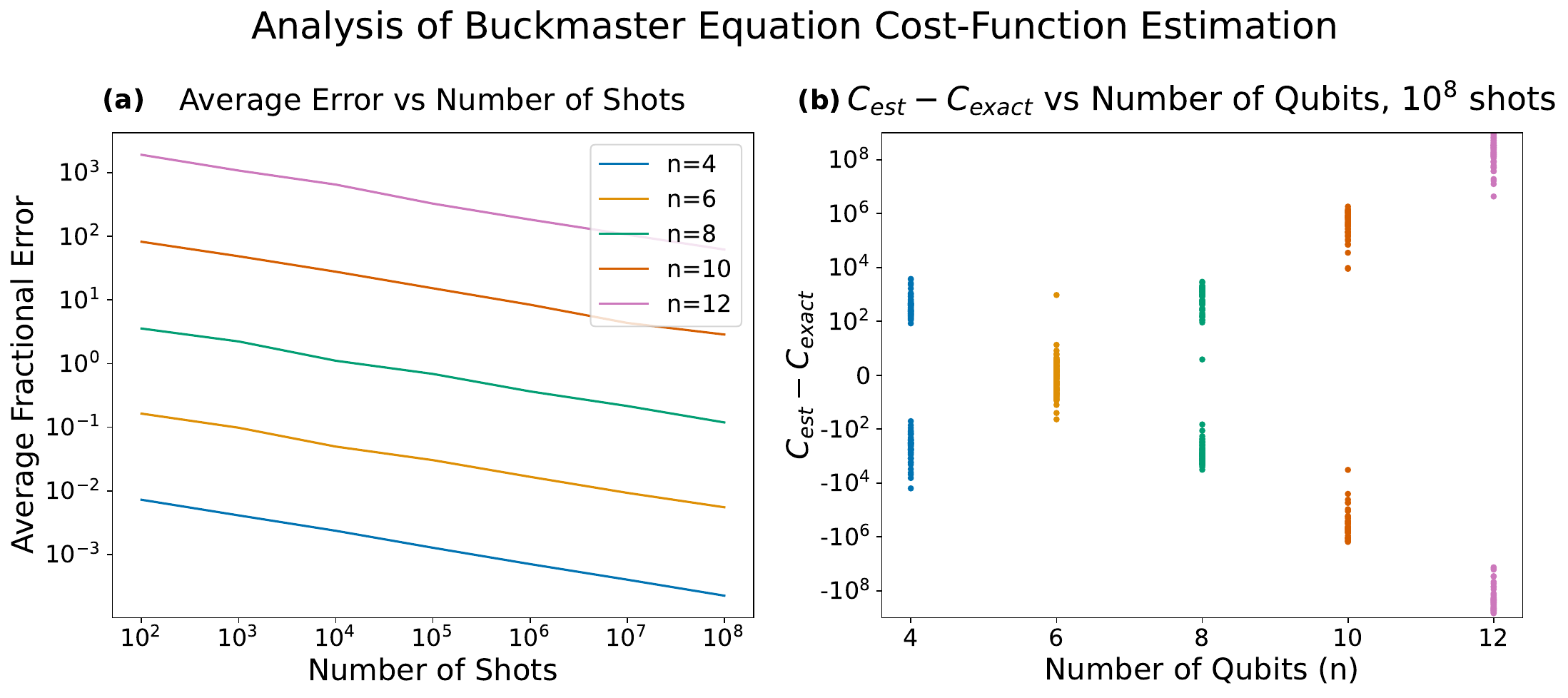}
     \hfill
    \caption{\textbf{Analysis of Buckmaster Cost-Function Estimation}. We estimated the cost-function by estimating each relevant expectation value using a binomial experiment with the given number of shots, effectively replicating what the output from a noiseless quantum computer would be. We repeat this experiment 100 times, to get 100 different estimates of the cost-function. a) Average fractional error, defined as the average of $|C_\textrm{est} - C_\textrm{exact}|/|C_\textrm{exact}|$ over all 100 trials as a function of shots. The slope of each line in the log-log plot is very close to -0.5, while the y-intercept is exponentially increasing with the number of qubits, exhibiting a dependence of $\textrm{error} \propto e^n\cdot \textrm{shots}^{-1/2}$. This suggests that to achieve an average error of $\epsilon$ in the estimation of the cost-function, one needs $O(\frac{e^n}{\epsilon^2})$ shots (ignoring the effect of noise from hardware imperfections). b) Scatter plot of all 100 samples of the cost-function with $10^8$ shots for each number of qubits. The estimations of the cost-function tend to exponentially deviate away from the true value as the number of qubits increases.}
    \label{fig:uncertainty}
\end{figure*}

An important consideration for the variational-PDE-solving method discussed in this paper is the precision with which the expectation values appearing in the cost functions must be estimated, which in turn impacts the computational cost of computing the cost function.\footnote{Ref.~\cite{Yangyang2022} presents a related study of the cost for solving linear PDEs using the variational-quantum-linear-solver approach.} The cost functions contain expectation values (whose magnitudes are $O(1)$) being multiplied by prefactors that are exponential in $n$; these prefactors arise from the finite-difference methods used to derive the cost functions. This suggests that one may have to estimate the expectation values to a precision that increases exponentially with $n$ in order to get a sufficiently accurate estimate of the cost function. Suppose we want to estimate the cost function with a maximum error of $\epsilon$. To do this, we need to estimate certain expectation values with an error of less than $O(\epsilon/e^{n})$. The error in the estimation of $\langle \mathcal{O} \rangle$ using $M$ shots of the Hadamard Test is given by $\mathcal{E}_{M}(\mathcal{O}) \sim \frac{\sqrt{\langle \mathcal{O}^{2} \rangle - \langle \mathcal{O} \rangle^{2}}}{\sqrt{M}}.$ This means that to get $\mathcal{E}_{M} = O(\epsilon/e^{n})$, $M$ must scale as $O(e^{2n}/\epsilon^{2})$. This is characterized numerically in Figure~\ref{fig:uncertainty}, in which we plot how the estimated value of the cost function for the Buckmaster equation converges to its true value as a function of the number of qubits and the number of shots, displaying the scaling behavior derived above. This issue is relevant even for relatively small problem sizes: for example, when solving the Buckmaster equation with $n=6$ qubits, expectation values must be sampled with $10^8$ shots to achieve around $1\%$ error in the estimation of the cost function, which we found to be roughly the minimum required accuracy for the optimization procedure to converge to the solution of the Buckmaster equation. The required number of shots to achieve acceptable accuracy constrains what can be demonstrated on real hardware (due to both the costs in money and time of hardware shots), and with the scaling of the present approach for formulating and estimating cost functions, ultimately presents an obstruction to achieving quantum advantage. This problem appears to persist to higher dimensional problems. For example, in the 2D linear BSE cost function (Appendix \ref{appendix:2DBSE}), there are terms with prefactors that scale as $O(e^{n_x+n_y}) = O(e^{n})$, suggesting that even for high dimensional problems with a few number of qubits per dimension, the exponential scaling in the total number of qubits is still present.


\begin{figure*}[t]
    \centering
    \includegraphics[width=17cm]{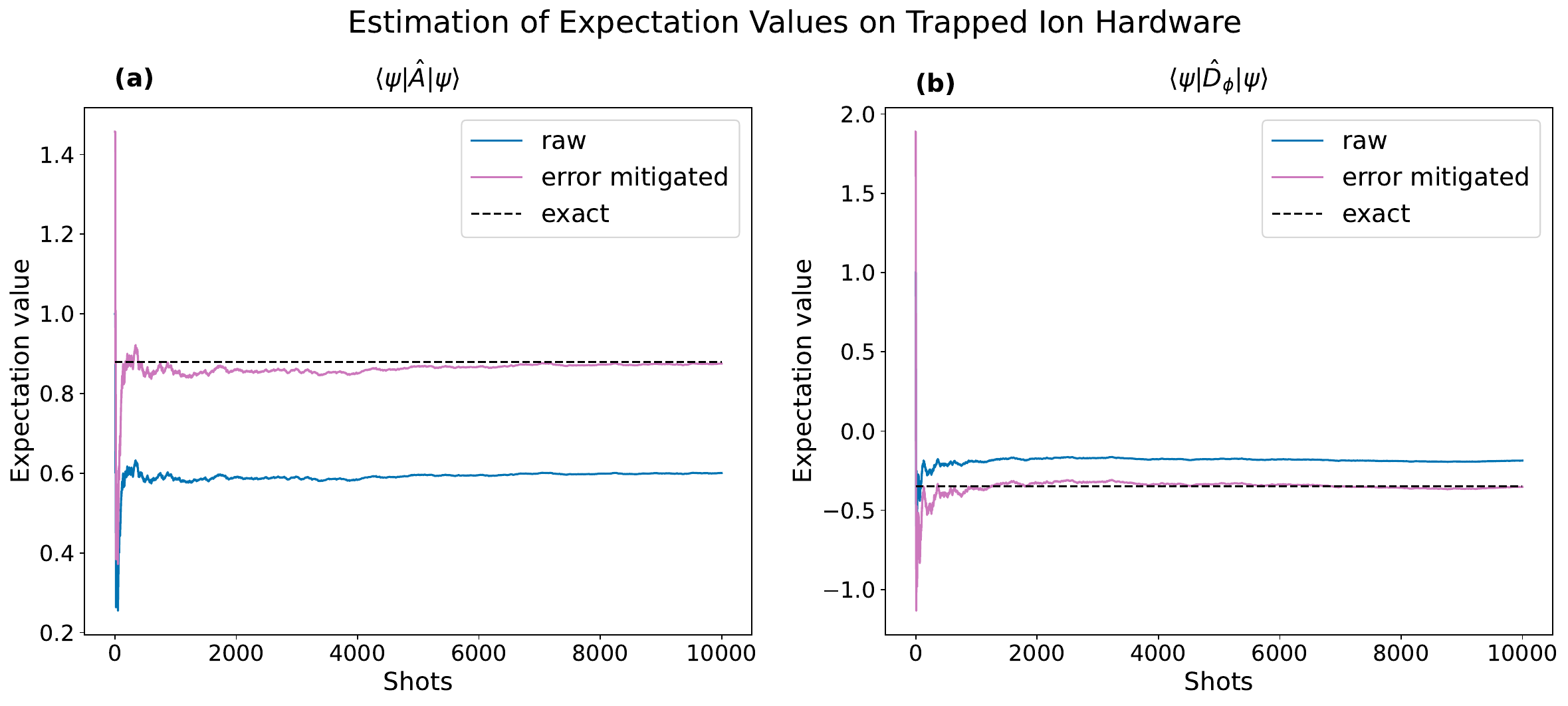}
    \caption{\textbf{Estimation of Expectation Values on a Trapped-Ion Processor}. This data was generated by runs on an 11-qubit trapped-ion device offered by IonQ \cite{IONQ}. The first expectation value was calculated with 2 ansatz qubits using the ULA. The second was calculated with 1 ansatz qubit, using a single $R_y$ gate as the ansatz. The circuits are shown in Appendix \ref{appendix:Circuits}. To mitigate the errors in the estimation of the expectation values, we ran each circuit with one set of parameters for which the expectation value is known, in order to estimate the damping factor \cite{Eliott} of the circuits. We then run the circuits with the true set of parameters, and divide by the damping factor to recover a better estimate of the true expectation value. Using this method, we find errors of 0.552\% and 1.56\% in the estimation of the first and second expectation value, respectively.}
    \label{fig:hardware}
\end{figure*}


We sought to demonstrate that the cost functions we derived to solve the nonlinear-Black--Scholes equation can be computed on prototype quantum hardware. To this end, we estimated two expectation values appearing in the cost function on a trapped-ion quantum processor from IonQ \cite{IONQ}, albeit in the admittedly toy case of having just $\leq 2$ ansatz qubits. In particular, using the Hadamard Test, we estimate oned ``linear" expectation value, $\braket{\psi|\hat{A}|\psi}$ with 2 ansatz qubits, and one ``nonlinear" expectation value, $\braket{\psi|\hat{D}_\phi|\psi}$ with 1 ansatz qubit, as a function of number of measurement shots. Both circuits had a total of 3 qubits. The results are shown in Figure~\ref{fig:hardware}; we were able to estimate both expectation values to within an accuracy of approximately 1\% using the error-mitigation technique described in Ref.~\cite{Eliott}. We were unable to estimate expectation values like $\braket{\psi|\tilde{\psi}}$ due to the controlled ansaetze appearing in the circuits for computing them, which require many more two-qubit gates than we could reliably run on the hardware available to us. This is not a situation specific to the nonlinear-Black--Scholes equation: in general, controlled ansaetze will appear in the circuits for computing some of the terms of the cost function for a nonlinear PDE, yielding rather deep circuits when more than a few qubit qubits are used (unless the ansatz is exceptionally simple, in which case it will probably also be insufficiently expressive).

\section{Conclusion and Outlook}
\label{sec:conclusion}
In this paper we presented an extension of the variational PDE-solving algorithm introduced in Ref.~\cite{lubasch}, and demonstrated how it could be applied to a wide class of nonlinear and multidimensional PDEs that are first order in time. We demonstrated that the algorithm is amenable to various explicit and semi-implicit Runge--Kutta numerical schemes, applying both the Forward Euler and Backward Euler discretizations in simulation. We used two ansaetze: a Fourier ansatz (the real-valued ZGR-QFT ansatz; Appendix~\ref{appendix:ZGRQFT}) and a hardware-efficient ansatz (the Universal Layered Ansatz; Appendix~\ref{appendix:ULA}), which is universal if sufficiently many layers are used. We benchmarked these ansatze for expressibility and trainability using several popular optimization algorithms for minimizing the cost-functions we derived. While both ansaetze are fully general, the ZGR-QFT is more efficient for representing smooth functions, while we found that the ULA can be more efficient for representing certain ``jagged'' functions. We then, in simulation, applied the quantum PDE solving method to solving a nonlinear Black--Scholes equation, the double-asset linear Black--Scholes equation, the Buckmaster equation, and the deterministic Kardar--Parisi--Zhang equation. We were able to achieve accurate solutions for up to $n=12$ ansatz qubits in noiseless simulations and demonstrate proof-of-concept results on a trapped-ion quantum computer with up to $n=2$ ansatz qubits. We did not apply the quantum PDE solving method to an equation which is both nonlinear and multidimensional due to the difficulty of simulating a large number of qubits (e.g. the 2D nonlinear BSE with $6+6$ ansatz qubits would require simulating circuits with over 36 qubits), though in principle the method is amenable to nonlinear multidimensional equations. During our investigations we encountered several challenges that appear necessary to address before one can hope to solve PDEs using quantum variational approaches in a regime that might give a quantum advantage over classical PDE solvers:

\begin{itemize}
    \item \textbf{Sufficiently accurate estimation of the value of the cost function.} The cost functions used in variational PDE solving involve expectation values (whose magnitudes are $O(1)$) multiplied by prefactors that are exponential in the number of qubits $n$. This suggests that the expectation values must be estimated with an exponentially increasing precision (in $n$) in order to calculate the cost functions accurately enough that the optimizer can steer towards good solutions rather than poor ones. If one uses a method (such as the Hadamard Test) requiring exponentially many circuit evaluations (shots) to achieve exponential precision, the approach both becomes impractical to test with moderately sized ($n>10$) ansaetze on current hardware and has scaling that likely rules out a quantum advantage.
    
    \item \textbf{Design of ansaetze that are sufficiently expressive, use few parameters, and don't render the algorithm classical simulable.} The ideal variational ansatz can probe the region of Hilbert space where the problem solutions lie with as few parameters as possible (e.g., needing only $n$ parameters would be very favorable) while also remaining classically intractable, in the sense that there should not be a polynomial-time classical algorithm to simulate circuits involving the ansatz. The desirable properties which make the ZGR-QFT good at representing smooth functions also make it classically tractable. It is an open problem to identify an ansatz that has the desirable properties (unchaotic, low number of parameters, adequately expressive, trainable) while remaining classically intractable, or, if none exists, to elucidate the tradeoffs that the best possible ansaetze have.
    
    \item \textbf{Deep circuits.} The circuits appearing in the variational-PDE-solving algorithm involve controlled applications of the ansatz unitary. If the ansatz uses many two-qubit gates, then the controlled ansatz will be difficult to execute on near-term hardware. In our demonstrations on IonQ's 11-qubit device, we found that the full algorithm for solving simple nonlinear PDEs is not practical to run on current hardware due to the circuit depths involved, even with as few as 2 ansatz qubits.
\end{itemize}

These findings are consistent with Ref.~\cite{Guseynov2022}'s analysis of quantum variational solving of a linear PDE (the heat equation). We hope that through alternative constructions of cost functions, or alternative methods of evaluating the cost functions, as well as the design of shallow ansaetze tailored to specific PDEs, that progress can be made toward practical nonlinear-PDE solving using variational quantum algorithms. 

Lastly, we acknowledge the recent work Ref.~\cite{lubasch_full_evo} in which the authors propose a variational PDE time evolution scheme in which the entire time evolution is captured in one optimization procedure by utilizing an additional qubit register corresponding to the time dimension. We find the authors' ideas to be interesting and promising. They argue that the temporal discretization can be chosen such that the exponential factors associated with the spatial and temporal discretizations cancel out, circumventing the issues described in Section \ref{sec:hardware}. However, such a scheme may prevent the use of clever ansatze such as the ZGR-QFT which take advantage of the spatial structure of candidate solutions, making cost function optimization much more difficult.

\section{Code and Data Availability}
\label{sec:code}

The code used to run the numerical simulations, the hardware runs on IonQ, and to generate the figures in this paper, as well as the data from our simulations/hardware demonstrations, is open source and available at \url{https://doi.org/10.5281/zenodo.10035285}. This includes implementations of the ZGR-QFT ansatz, the Universal Layered Ansatz, and the cost functions for the PDEs considered in the paper.

\section{Author Contributions}
\label{sec:contributions}

P.L.M., A.S., and T.W.W. conceived the project. A.S. derived the cost functions, coded the time-evolution circuits, ran the numerical simulations and hardware demonstrations, and analyzed the data. Y.L. and M.M. helped derive the cost functions, Y.L. helped debug the code for the circuits, and M.M. helped run the simulations. M.M. and T.W.W. developed and provided the code for the 1D and 2D ZGR-QFT ansatz, and wrote Appendix~\ref{appendix:ZGRQFT}. A.S. and P.L.M. wrote the main manuscript, revised by all authors. P.L.M. supervised the project.


\bibliography{references}

\begin{thebibliography}{36}%
\makeatletter
\providecommand \@ifxundefined [1]{%
 \@ifx{#1\undefined}
}%
\providecommand \@ifnum [1]{%
 \ifnum #1\expandafter \@firstoftwo
 \else \expandafter \@secondoftwo
 \fi
}%
\providecommand \@ifx [1]{%
 \ifx #1\expandafter \@firstoftwo
 \else \expandafter \@secondoftwo
 \fi
}%
\providecommand \natexlab [1]{#1}%
\providecommand \enquote  [1]{``#1''}%
\providecommand \bibnamefont  [1]{#1}%
\providecommand \bibfnamefont [1]{#1}%
\providecommand \citenamefont [1]{#1}%
\providecommand \href@noop [0]{\@secondoftwo}%
\providecommand \href [0]{\begingroup \@sanitize@url \@href}%
\providecommand \@href[1]{\@@startlink{#1}\@@href}%
\providecommand \@@href[1]{\endgroup#1\@@endlink}%
\providecommand \@sanitize@url [0]{\catcode `\\12\catcode `\$12\catcode `\&12\catcode `\#12\catcode `\^12\catcode `\_12\catcode `\%12\relax}%
\providecommand \@@startlink[1]{}%
\providecommand \@@endlink[0]{}%
\providecommand \url  [0]{\begingroup\@sanitize@url \@url }%
\providecommand \@url [1]{\endgroup\@href {#1}{\urlprefix }}%
\providecommand \urlprefix  [0]{URL }%
\providecommand \Eprint [0]{\href }%
\providecommand \doibase [0]{https://doi.org/}%
\providecommand \selectlanguage [0]{\@gobble}%
\providecommand \bibinfo  [0]{\@secondoftwo}%
\providecommand \bibfield  [0]{\@secondoftwo}%
\providecommand \translation [1]{[#1]}%
\providecommand \BibitemOpen [0]{}%
\providecommand \bibitemStop [0]{}%
\providecommand \bibitemNoStop [0]{.\EOS\space}%
\providecommand \EOS [0]{\spacefactor3000\relax}%
\providecommand \BibitemShut  [1]{\csname bibitem#1\endcsname}%
\let\auto@bib@innerbib\@empty
\bibitem [{\citenamefont {{Lubasch}}\ \emph {et~al.}(2020)\citenamefont {{Lubasch}}, \citenamefont {{Joo}}, \citenamefont {{Moinier}}, \citenamefont {{Kiffner}},\ and\ \citenamefont {{Jaksch}}}]{lubasch}%
  \BibitemOpen
  \bibfield  {author} {\bibinfo {author} {\bibfnamefont {M.}~\bibnamefont {{Lubasch}}}, \bibinfo {author} {\bibfnamefont {J.}~\bibnamefont {{Joo}}}, \bibinfo {author} {\bibfnamefont {P.}~\bibnamefont {{Moinier}}}, \bibinfo {author} {\bibfnamefont {M.}~\bibnamefont {{Kiffner}}},\ and\ \bibinfo {author} {\bibfnamefont {D.}~\bibnamefont {{Jaksch}}},\ }\bibfield  {title} {\bibinfo {title} {{Variational quantum algorithms for nonlinear problems}},\ }\href {https://doi.org/10.1103/PhysRevA.101.010301} {\bibfield  {journal} {\bibinfo  {journal} {\pra}\ }\textbf {\bibinfo {volume} {101}},\ \bibinfo {eid} {010301(R)} (\bibinfo {year} {2020})},\ \Eprint {https://arxiv.org/abs/1907.09032} {arXiv:1907.09032 [quant-ph]} \BibitemShut {NoStop}%
\bibitem [{\citenamefont {{Wright}}\ \emph {et~al.}(2019)\citenamefont {{Wright}}, \citenamefont {{Beck}}, \citenamefont {{Debnath}}, \citenamefont {{Amini}}, \citenamefont {{Nam}}, \citenamefont {{Grzesiak}}, \citenamefont {{Chen}}, \citenamefont {{Pisenti}}, \citenamefont {{Chmielewski}}, \citenamefont {{Collins}}, \citenamefont {{Hudek}}, \citenamefont {{Mizrahi}}, \citenamefont {{Wong-Campos}}, \citenamefont {{Allen}}, \citenamefont {{Apisdorf}}, \citenamefont {{Solomon}}, \citenamefont {{Williams}}, \citenamefont {{Ducore}}, \citenamefont {{Blinov}}, \citenamefont {{Kreikemeier}}, \citenamefont {{Chaplin}}, \citenamefont {{Keesan}}, \citenamefont {{Monroe}},\ and\ \citenamefont {{Kim}}}]{IONQ}%
  \BibitemOpen
  \bibfield  {author} {\bibinfo {author} {\bibfnamefont {K.}~\bibnamefont {{Wright}}}, \bibinfo {author} {\bibfnamefont {K.~M.}\ \bibnamefont {{Beck}}}, \bibinfo {author} {\bibfnamefont {S.}~\bibnamefont {{Debnath}}}, \bibinfo {author} {\bibfnamefont {J.~M.}\ \bibnamefont {{Amini}}}, \bibinfo {author} {\bibfnamefont {Y.}~\bibnamefont {{Nam}}}, \bibinfo {author} {\bibfnamefont {N.}~\bibnamefont {{Grzesiak}}}, \bibinfo {author} {\bibfnamefont {J.~S.}\ \bibnamefont {{Chen}}}, \bibinfo {author} {\bibfnamefont {N.~C.}\ \bibnamefont {{Pisenti}}}, \bibinfo {author} {\bibfnamefont {M.}~\bibnamefont {{Chmielewski}}}, \bibinfo {author} {\bibfnamefont {C.}~\bibnamefont {{Collins}}}, \bibinfo {author} {\bibfnamefont {K.~M.}\ \bibnamefont {{Hudek}}}, \bibinfo {author} {\bibfnamefont {J.}~\bibnamefont {{Mizrahi}}}, \bibinfo {author} {\bibfnamefont {J.~D.}\ \bibnamefont {{Wong-Campos}}}, \bibinfo {author} {\bibfnamefont {S.}~\bibnamefont {{Allen}}}, \bibinfo {author} {\bibfnamefont {J.}~\bibnamefont {{Apisdorf}}}, \bibinfo
  {author} {\bibfnamefont {P.}~\bibnamefont {{Solomon}}}, \bibinfo {author} {\bibfnamefont {M.}~\bibnamefont {{Williams}}}, \bibinfo {author} {\bibfnamefont {A.~M.}\ \bibnamefont {{Ducore}}}, \bibinfo {author} {\bibfnamefont {A.}~\bibnamefont {{Blinov}}}, \bibinfo {author} {\bibfnamefont {S.~M.}\ \bibnamefont {{Kreikemeier}}}, \bibinfo {author} {\bibfnamefont {V.}~\bibnamefont {{Chaplin}}}, \bibinfo {author} {\bibfnamefont {M.}~\bibnamefont {{Keesan}}}, \bibinfo {author} {\bibfnamefont {C.}~\bibnamefont {{Monroe}}},\ and\ \bibinfo {author} {\bibfnamefont {J.}~\bibnamefont {{Kim}}},\ }\bibfield  {title} {\bibinfo {title} {{Benchmarking an 11-qubit quantum computer}},\ }\href {https://doi.org/10.1038/s41467-019-13534-2} {\bibfield  {journal} {\bibinfo  {journal} {Nature Communications}\ }\textbf {\bibinfo {volume} {10}},\ \bibinfo {eid} {5464} (\bibinfo {year} {2019})},\ \Eprint {https://arxiv.org/abs/1903.08181} {arXiv:1903.08181 [quant-ph]} \BibitemShut {NoStop}%
\bibitem [{\citenamefont {{Cerezo}}\ \emph {et~al.}(2021)\citenamefont {{Cerezo}}, \citenamefont {{Arrasmith}}, \citenamefont {{Babbush}}, \citenamefont {{Benjamin}}, \citenamefont {{Endo}}, \citenamefont {{Fujii}}, \citenamefont {{McClean}}, \citenamefont {{Mitarai}}, \citenamefont {{Yuan}}, \citenamefont {{Cincio}},\ and\ \citenamefont {{Coles}}}]{Cerezo2020-1}%
  \BibitemOpen
  \bibfield  {author} {\bibinfo {author} {\bibfnamefont {M.}~\bibnamefont {{Cerezo}}}, \bibinfo {author} {\bibfnamefont {A.}~\bibnamefont {{Arrasmith}}}, \bibinfo {author} {\bibfnamefont {R.}~\bibnamefont {{Babbush}}}, \bibinfo {author} {\bibfnamefont {S.~C.}\ \bibnamefont {{Benjamin}}}, \bibinfo {author} {\bibfnamefont {S.}~\bibnamefont {{Endo}}}, \bibinfo {author} {\bibfnamefont {K.}~\bibnamefont {{Fujii}}}, \bibinfo {author} {\bibfnamefont {J.~R.}\ \bibnamefont {{McClean}}}, \bibinfo {author} {\bibfnamefont {K.}~\bibnamefont {{Mitarai}}}, \bibinfo {author} {\bibfnamefont {X.}~\bibnamefont {{Yuan}}}, \bibinfo {author} {\bibfnamefont {L.}~\bibnamefont {{Cincio}}},\ and\ \bibinfo {author} {\bibfnamefont {P.~J.}\ \bibnamefont {{Coles}}},\ }\bibfield  {title} {\bibinfo {title} {{Variational Quantum Algorithms}},\ }\href@noop {} {\bibfield  {journal} {\bibinfo  {journal} {Nature Reviews Physics}\ }\textbf {\bibinfo {volume} {3}},\ \bibinfo {pages} {625} (\bibinfo {year} {2021})}\BibitemShut {NoStop}%
\bibitem [{\citenamefont {Nielsen}\ and\ \citenamefont {Chuang}(2010)}]{nielsen2010quantum}%
  \BibitemOpen
  \bibfield  {author} {\bibinfo {author} {\bibfnamefont {M.~A.}\ \bibnamefont {Nielsen}}\ and\ \bibinfo {author} {\bibfnamefont {I.~L.}\ \bibnamefont {Chuang}},\ }\href@noop {} {\emph {\bibinfo {title} {Quantum computation and quantum information}}}\ (\bibinfo  {publisher} {Cambridge University Press},\ \bibinfo {year} {2010})\BibitemShut {NoStop}%
\bibitem [{\citenamefont {Preskill}(2018)}]{preskill2018quantum}%
  \BibitemOpen
  \bibfield  {author} {\bibinfo {author} {\bibfnamefont {J.}~\bibnamefont {Preskill}},\ }\bibfield  {title} {\bibinfo {title} {Quantum computing in the nisq era and beyond},\ }\href@noop {} {\bibfield  {journal} {\bibinfo  {journal} {Quantum}\ }\textbf {\bibinfo {volume} {2}},\ \bibinfo {pages} {79} (\bibinfo {year} {2018})}\BibitemShut {NoStop}%
\bibitem [{\citenamefont {{Kyriienko}}\ \emph {et~al.}(2020)\citenamefont {{Kyriienko}}, \citenamefont {{Paine}},\ and\ \citenamefont {{Elfving}}}]{Kyriienko2020}%
  \BibitemOpen
  \bibfield  {author} {\bibinfo {author} {\bibfnamefont {O.}~\bibnamefont {{Kyriienko}}}, \bibinfo {author} {\bibfnamefont {A.~E.}\ \bibnamefont {{Paine}}},\ and\ \bibinfo {author} {\bibfnamefont {V.~E.}\ \bibnamefont {{Elfving}}},\ }\bibfield  {title} {\bibinfo {title} {{Solving nonlinear differential equations with differentiable quantum circuits}},\ }\href@noop {} {\bibfield  {journal} {\bibinfo  {journal} {arXiv e-prints}\ ,\ \bibinfo {eid} {arXiv:2011.10395}} (\bibinfo {year} {2020})},\ \Eprint {https://arxiv.org/abs/2011.10395} {arXiv:2011.10395 [quant-ph]} \BibitemShut {NoStop}%
\bibitem [{\citenamefont {Liu}\ \emph {et~al.}(2021)\citenamefont {Liu}, \citenamefont {Wu}, \citenamefont {Wan}, \citenamefont {Pan}, \citenamefont {Qin}, \citenamefont {Gao},\ and\ \citenamefont {Wen}}]{liu2021variational}%
  \BibitemOpen
  \bibfield  {author} {\bibinfo {author} {\bibfnamefont {H.-L.}\ \bibnamefont {Liu}}, \bibinfo {author} {\bibfnamefont {Y.-S.}\ \bibnamefont {Wu}}, \bibinfo {author} {\bibfnamefont {L.-C.}\ \bibnamefont {Wan}}, \bibinfo {author} {\bibfnamefont {S.-J.}\ \bibnamefont {Pan}}, \bibinfo {author} {\bibfnamefont {S.-J.}\ \bibnamefont {Qin}}, \bibinfo {author} {\bibfnamefont {F.}~\bibnamefont {Gao}},\ and\ \bibinfo {author} {\bibfnamefont {Q.-Y.}\ \bibnamefont {Wen}},\ }\bibfield  {title} {\bibinfo {title} {Variational quantum algorithm for the {P}oisson equation},\ }\href@noop {} {\bibfield  {journal} {\bibinfo  {journal} {Physical Review A}\ }\textbf {\bibinfo {volume} {104}},\ \bibinfo {pages} {022418} (\bibinfo {year} {2021})}\BibitemShut {NoStop}%
\bibitem [{\citenamefont {{Mocz}}\ and\ \citenamefont {{Szasz}}(2021)}]{Mocz2021}%
  \BibitemOpen
  \bibfield  {author} {\bibinfo {author} {\bibfnamefont {P.}~\bibnamefont {{Mocz}}}\ and\ \bibinfo {author} {\bibfnamefont {A.}~\bibnamefont {{Szasz}}},\ }\bibfield  {title} {\bibinfo {title} {{Toward Cosmological Simulations of Dark Matter on Quantum Computers}},\ }\href {https://doi.org/10.3847/1538-4357/abe6ac} {\bibfield  {journal} {\bibinfo  {journal} {\apj}\ }\textbf {\bibinfo {volume} {910}},\ \bibinfo {eid} {29} (\bibinfo {year} {2021})},\ \Eprint {https://arxiv.org/abs/2101.05821} {arXiv:2101.05821 [astro-ph.CO]} \BibitemShut {NoStop}%
\bibitem [{\citenamefont {{Sato}}\ \emph {et~al.}(2021)\citenamefont {{Sato}}, \citenamefont {{Kondo}}, \citenamefont {{Koide}}, \citenamefont {{Takamatsu}},\ and\ \citenamefont {{Imoto}}}]{Sato2021}%
  \BibitemOpen
  \bibfield  {author} {\bibinfo {author} {\bibfnamefont {Y.}~\bibnamefont {{Sato}}}, \bibinfo {author} {\bibfnamefont {R.}~\bibnamefont {{Kondo}}}, \bibinfo {author} {\bibfnamefont {S.}~\bibnamefont {{Koide}}}, \bibinfo {author} {\bibfnamefont {H.}~\bibnamefont {{Takamatsu}}},\ and\ \bibinfo {author} {\bibfnamefont {N.}~\bibnamefont {{Imoto}}},\ }\bibfield  {title} {\bibinfo {title} {{Variational quantum algorithm based on the minimum potential energy for solving the Poisson equation}},\ }\href {https://doi.org/10.1103/PhysRevA.104.052409} {\bibfield  {journal} {\bibinfo  {journal} {\pra}\ }\textbf {\bibinfo {volume} {104}},\ \bibinfo {eid} {052409} (\bibinfo {year} {2021})},\ \Eprint {https://arxiv.org/abs/2106.09333} {arXiv:2106.09333 [quant-ph]} \BibitemShut {NoStop}%
\bibitem [{\citenamefont {Garc{\'\i}a-Molina}\ \emph {et~al.}(2022)\citenamefont {Garc{\'\i}a-Molina}, \citenamefont {Rodr{\'\i}guez-Mediavilla},\ and\ \citenamefont {Garc{\'\i}a-Ripoll}}]{garcia2022quantum}%
  \BibitemOpen
  \bibfield  {author} {\bibinfo {author} {\bibfnamefont {P.}~\bibnamefont {Garc{\'\i}a-Molina}}, \bibinfo {author} {\bibfnamefont {J.}~\bibnamefont {Rodr{\'\i}guez-Mediavilla}},\ and\ \bibinfo {author} {\bibfnamefont {J.~J.}\ \bibnamefont {Garc{\'\i}a-Ripoll}},\ }\bibfield  {title} {\bibinfo {title} {Quantum fourier analysis for multivariate functions and applications to a class of schr{\"o}dinger-type partial differential equations},\ }\href@noop {} {\bibfield  {journal} {\bibinfo  {journal} {Physical Review A}\ }\textbf {\bibinfo {volume} {105}},\ \bibinfo {pages} {012433} (\bibinfo {year} {2022})}\BibitemShut {NoStop}%
\bibitem [{\citenamefont {{Yew Leong}}\ \emph {et~al.}(2022)\citenamefont {{Yew Leong}}, \citenamefont {{Ewe}},\ and\ \citenamefont {{Enshan Koh}}}]{YewLeong2022}%
  \BibitemOpen
  \bibfield  {author} {\bibinfo {author} {\bibfnamefont {F.}~\bibnamefont {{Yew Leong}}}, \bibinfo {author} {\bibfnamefont {W.-B.}\ \bibnamefont {{Ewe}}},\ and\ \bibinfo {author} {\bibfnamefont {D.}~\bibnamefont {{Enshan Koh}}},\ }\bibfield  {title} {\bibinfo {title} {{Variational Quantum Evolution Equation Solver}},\ }\href@noop {} {\bibfield  {journal} {\bibinfo  {journal} {arXiv e-prints}\ ,\ \bibinfo {eid} {arXiv:2204.02912}} (\bibinfo {year} {2022})},\ \Eprint {https://arxiv.org/abs/2204.02912} {arXiv:2204.02912 [quant-ph]} \BibitemShut {NoStop}%
\bibitem [{\citenamefont {{Yew Leong}}\ \emph {et~al.}(2023)\citenamefont {{Yew Leong}}, \citenamefont {{Enshan Koh}}, \citenamefont {Ewe},\ and\ \citenamefont {Kong}}]{YewLeong2023}%
  \BibitemOpen
  \bibfield  {author} {\bibinfo {author} {\bibfnamefont {F.}~\bibnamefont {{Yew Leong}}}, \bibinfo {author} {\bibfnamefont {D.}~\bibnamefont {{Enshan Koh}}}, \bibinfo {author} {\bibfnamefont {W.-B.}\ \bibnamefont {Ewe}},\ and\ \bibinfo {author} {\bibfnamefont {J.~F.}\ \bibnamefont {Kong}},\ }\href@noop {} {\bibinfo {title} {Variational quantum simulation of partial differential equations: Applications in colloidal transport}} (\bibinfo {year} {2023}),\ \Eprint {https://arxiv.org/abs/2307.07173} {arXiv:2307.07173 [quant-ph]} \BibitemShut {NoStop}%
\bibitem [{\citenamefont {Ewe}\ \emph {et~al.}(2022)\citenamefont {Ewe}, \citenamefont {{Enshan Koh}}, \citenamefont {Goh}, \citenamefont {Chu},\ and\ \citenamefont {Png}}]{Ewe2022}%
  \BibitemOpen
  \bibfield  {author} {\bibinfo {author} {\bibfnamefont {W.-B.}\ \bibnamefont {Ewe}}, \bibinfo {author} {\bibfnamefont {D.}~\bibnamefont {{Enshan Koh}}}, \bibinfo {author} {\bibfnamefont {S.~T.}\ \bibnamefont {Goh}}, \bibinfo {author} {\bibfnamefont {H.-S.}\ \bibnamefont {Chu}},\ and\ \bibinfo {author} {\bibfnamefont {C.~E.}\ \bibnamefont {Png}},\ }\bibfield  {title} {\bibinfo {title} {Variational quantum-based simulation of waveguide modes},\ }\href {https://doi.org/10.1109/tmtt.2022.3151510} {\bibfield  {journal} {\bibinfo  {journal} {{IEEE} Transactions on Microwave Theory and Techniques}\ }\textbf {\bibinfo {volume} {70}},\ \bibinfo {pages} {2517} (\bibinfo {year} {2022})}\BibitemShut {NoStop}%
\bibitem [{\citenamefont {{Mouton}}\ \emph {et~al.}(2023)\citenamefont {{Mouton}}, \citenamefont {{Reiter}}, \citenamefont {{Chen}},\ and\ \citenamefont {{Rebentrost}}}]{Rebentrost2023}%
  \BibitemOpen
  \bibfield  {author} {\bibinfo {author} {\bibfnamefont {L.}~\bibnamefont {{Mouton}}}, \bibinfo {author} {\bibfnamefont {F.}~\bibnamefont {{Reiter}}}, \bibinfo {author} {\bibfnamefont {Y.}~\bibnamefont {{Chen}}},\ and\ \bibinfo {author} {\bibfnamefont {P.}~\bibnamefont {{Rebentrost}}},\ }\bibfield  {title} {\bibinfo {title} {{Towards Deep Learning-Based Quantum Algorithms for Solving Nonlinear Partial Differential Equations}},\ }\href {https://doi.org/10.48550/arXiv.2305.02019} {\bibfield  {journal} {\bibinfo  {journal} {arXiv e-prints}\ ,\ \bibinfo {eid} {arXiv:2305.02019}} (\bibinfo {year} {2023})},\ \Eprint {https://arxiv.org/abs/2305.02019} {arXiv:2305.02019 [quant-ph]} \BibitemShut {NoStop}%
\bibitem [{\citenamefont {{Amaro}}\ and\ \citenamefont {{Cruz}}(2023)}]{Amaro2023}%
  \BibitemOpen
  \bibfield  {author} {\bibinfo {author} {\bibfnamefont {{\'O}.}~\bibnamefont {{Amaro}}}\ and\ \bibinfo {author} {\bibfnamefont {D.}~\bibnamefont {{Cruz}}},\ }\bibfield  {title} {\bibinfo {title} {{A Living Review of Quantum Computing for Plasma Physics}},\ }\href {https://doi.org/10.48550/arXiv.2302.00001} {\bibfield  {journal} {\bibinfo  {journal} {arXiv e-prints}\ ,\ \bibinfo {eid} {arXiv:2302.00001}} (\bibinfo {year} {2023})},\ \Eprint {https://arxiv.org/abs/2302.00001} {arXiv:2302.00001 [physics.plasm-ph]} \BibitemShut {NoStop}%
\bibitem [{\citenamefont {Clader}\ \emph {et~al.}(2013)\citenamefont {Clader}, \citenamefont {Jacobs},\ and\ \citenamefont {Sprouse}}]{clader2013preconditioned}%
  \BibitemOpen
  \bibfield  {author} {\bibinfo {author} {\bibfnamefont {B.~D.}\ \bibnamefont {Clader}}, \bibinfo {author} {\bibfnamefont {B.~C.}\ \bibnamefont {Jacobs}},\ and\ \bibinfo {author} {\bibfnamefont {C.~R.}\ \bibnamefont {Sprouse}},\ }\bibfield  {title} {\bibinfo {title} {Preconditioned quantum linear system algorithm},\ }\href@noop {} {\bibfield  {journal} {\bibinfo  {journal} {Physical Review Letters}\ }\textbf {\bibinfo {volume} {110}},\ \bibinfo {pages} {250504} (\bibinfo {year} {2013})}\BibitemShut {NoStop}%
\bibitem [{\citenamefont {Cao}\ \emph {et~al.}(2013)\citenamefont {Cao}, \citenamefont {Papageorgiou}, \citenamefont {Petras}, \citenamefont {Traub},\ and\ \citenamefont {Kais}}]{cao2013quantum}%
  \BibitemOpen
  \bibfield  {author} {\bibinfo {author} {\bibfnamefont {Y.}~\bibnamefont {Cao}}, \bibinfo {author} {\bibfnamefont {A.}~\bibnamefont {Papageorgiou}}, \bibinfo {author} {\bibfnamefont {I.}~\bibnamefont {Petras}}, \bibinfo {author} {\bibfnamefont {J.}~\bibnamefont {Traub}},\ and\ \bibinfo {author} {\bibfnamefont {S.}~\bibnamefont {Kais}},\ }\bibfield  {title} {\bibinfo {title} {Quantum algorithm and circuit design solving the poisson equation},\ }\href@noop {} {\bibfield  {journal} {\bibinfo  {journal} {New Journal of Physics}\ }\textbf {\bibinfo {volume} {15}},\ \bibinfo {pages} {013021} (\bibinfo {year} {2013})}\BibitemShut {NoStop}%
\bibitem [{\citenamefont {Montanaro}\ and\ \citenamefont {Pallister}(2016)}]{montanaro2016quantum}%
  \BibitemOpen
  \bibfield  {author} {\bibinfo {author} {\bibfnamefont {A.}~\bibnamefont {Montanaro}}\ and\ \bibinfo {author} {\bibfnamefont {S.}~\bibnamefont {Pallister}},\ }\bibfield  {title} {\bibinfo {title} {Quantum algorithms and the finite element method},\ }\href@noop {} {\bibfield  {journal} {\bibinfo  {journal} {Physical Review A}\ }\textbf {\bibinfo {volume} {93}},\ \bibinfo {pages} {032324} (\bibinfo {year} {2016})}\BibitemShut {NoStop}%
\bibitem [{\citenamefont {{Arrazola}}\ \emph {et~al.}(2019)\citenamefont {{Arrazola}}, \citenamefont {{Kalajdzievski}}, \citenamefont {{Weedbrook}},\ and\ \citenamefont {{Lloyd}}}]{Arrazola2019}%
  \BibitemOpen
  \bibfield  {author} {\bibinfo {author} {\bibfnamefont {J.~M.}\ \bibnamefont {{Arrazola}}}, \bibinfo {author} {\bibfnamefont {T.}~\bibnamefont {{Kalajdzievski}}}, \bibinfo {author} {\bibfnamefont {C.}~\bibnamefont {{Weedbrook}}},\ and\ \bibinfo {author} {\bibfnamefont {S.}~\bibnamefont {{Lloyd}}},\ }\bibfield  {title} {\bibinfo {title} {{Quantum algorithm for nonhomogeneous linear partial differential equations}},\ }\href {https://doi.org/10.1103/PhysRevA.100.032306} {\bibfield  {journal} {\bibinfo  {journal} {\pra}\ }\textbf {\bibinfo {volume} {100}},\ \bibinfo {eid} {032306} (\bibinfo {year} {2019})},\ \Eprint {https://arxiv.org/abs/1809.02622} {arXiv:1809.02622 [quant-ph]} \BibitemShut {NoStop}%
\bibitem [{\citenamefont {{Childs}}\ \emph {et~al.}(2021)\citenamefont {{Childs}}, \citenamefont {{Liu}},\ and\ \citenamefont {{Ostrander}}}]{Childs2021}%
  \BibitemOpen
  \bibfield  {author} {\bibinfo {author} {\bibfnamefont {A.~M.}\ \bibnamefont {{Childs}}}, \bibinfo {author} {\bibfnamefont {J.-P.}\ \bibnamefont {{Liu}}},\ and\ \bibinfo {author} {\bibfnamefont {A.}~\bibnamefont {{Ostrander}}},\ }\bibfield  {title} {\bibinfo {title} {{High-precision quantum algorithms for partial differential equations}},\ }\href {https://doi.org/10.22331/q-2021-11-10-574} {\bibfield  {journal} {\bibinfo  {journal} {Quantum}\ }\textbf {\bibinfo {volume} {5}},\ \bibinfo {pages} {574} (\bibinfo {year} {2021})},\ \Eprint {https://arxiv.org/abs/2002.07868} {arXiv:2002.07868 [quant-ph]} \BibitemShut {NoStop}%
\bibitem [{\citenamefont {{Jin}}\ \emph {et~al.}(2022)\citenamefont {{Jin}}, \citenamefont {{Liu}},\ and\ \citenamefont {{Yu}}}]{Jin2022}%
  \BibitemOpen
  \bibfield  {author} {\bibinfo {author} {\bibfnamefont {S.}~\bibnamefont {{Jin}}}, \bibinfo {author} {\bibfnamefont {N.}~\bibnamefont {{Liu}}},\ and\ \bibinfo {author} {\bibfnamefont {Y.}~\bibnamefont {{Yu}}},\ }\bibfield  {title} {\bibinfo {title} {{Quantum simulation of partial differential equations via Schrodingerisation: technical details}},\ }\href {https://doi.org/10.48550/arXiv.2212.14703} {\bibfield  {journal} {\bibinfo  {journal} {arXiv e-prints}\ ,\ \bibinfo {eid} {arXiv:2212.14703}} (\bibinfo {year} {2022})},\ \Eprint {https://arxiv.org/abs/2212.14703} {arXiv:2212.14703 [quant-ph]} \BibitemShut {NoStop}%
\bibitem [{\citenamefont {Harrow}\ \emph {et~al.}(2009)\citenamefont {Harrow}, \citenamefont {Hassidim},\ and\ \citenamefont {Lloyd}}]{harrow2009quantum}%
  \BibitemOpen
  \bibfield  {author} {\bibinfo {author} {\bibfnamefont {A.~W.}\ \bibnamefont {Harrow}}, \bibinfo {author} {\bibfnamefont {A.}~\bibnamefont {Hassidim}},\ and\ \bibinfo {author} {\bibfnamefont {S.}~\bibnamefont {Lloyd}},\ }\bibfield  {title} {\bibinfo {title} {Quantum algorithm for linear systems of equations},\ }\href@noop {} {\bibfield  {journal} {\bibinfo  {journal} {Physical Review Letters}\ }\textbf {\bibinfo {volume} {103}},\ \bibinfo {pages} {150502} (\bibinfo {year} {2009})}\BibitemShut {NoStop}%
\bibitem [{\citenamefont {{Gonzalez-Conde}}\ \emph {et~al.}(2021)\citenamefont {{Gonzalez-Conde}}, \citenamefont {{Rodr{\'\i}guez-Rozas}}, \citenamefont {{Solano}},\ and\ \citenamefont {{Sanz}}}]{Gonzalez-Conde2021}%
  \BibitemOpen
  \bibfield  {author} {\bibinfo {author} {\bibfnamefont {J.}~\bibnamefont {{Gonzalez-Conde}}}, \bibinfo {author} {\bibfnamefont {{\'A}.}~\bibnamefont {{Rodr{\'\i}guez-Rozas}}}, \bibinfo {author} {\bibfnamefont {E.}~\bibnamefont {{Solano}}},\ and\ \bibinfo {author} {\bibfnamefont {M.}~\bibnamefont {{Sanz}}},\ }\bibfield  {title} {\bibinfo {title} {{Simulating option price dynamics with exponential quantum speedup}},\ }\href@noop {} {\bibfield  {journal} {\bibinfo  {journal} {arXiv e-prints}\ ,\ \bibinfo {eid} {arXiv:2101.04023}} (\bibinfo {year} {2021})},\ \Eprint {https://arxiv.org/abs/2101.04023} {arXiv:2101.04023 [quant-ph]} \BibitemShut {NoStop}%
\bibitem [{\citenamefont {{Cerezo}}\ \emph {et~al.}(2020)\citenamefont {{Cerezo}}, \citenamefont {{Sharma}}, \citenamefont {{Arrasmith}},\ and\ \citenamefont {{Coles}}}]{Cerezo2020-2}%
  \BibitemOpen
  \bibfield  {author} {\bibinfo {author} {\bibfnamefont {M.}~\bibnamefont {{Cerezo}}}, \bibinfo {author} {\bibfnamefont {K.}~\bibnamefont {{Sharma}}}, \bibinfo {author} {\bibfnamefont {A.}~\bibnamefont {{Arrasmith}}},\ and\ \bibinfo {author} {\bibfnamefont {P.~J.}\ \bibnamefont {{Coles}}},\ }\bibfield  {title} {\bibinfo {title} {{Variational Quantum State Eigensolver}},\ }\href@noop {} {\bibfield  {journal} {\bibinfo  {journal} {arXiv e-prints}\ ,\ \bibinfo {eid} {arXiv:2004.01372}} (\bibinfo {year} {2020})},\ \Eprint {https://arxiv.org/abs/2004.01372} {arXiv:2004.01372 [quant-ph]} \BibitemShut {NoStop}%
\bibitem [{\citenamefont {{Moosa}}\ \emph {et~al.}(2023)\citenamefont {{Moosa}}, \citenamefont {{Watts}}, \citenamefont {{Chen}}, \citenamefont {{Sarma}},\ and\ \citenamefont {{McMahon}}}]{Watts2023}%
  \BibitemOpen
  \bibfield  {author} {\bibinfo {author} {\bibfnamefont {M.}~\bibnamefont {{Moosa}}}, \bibinfo {author} {\bibfnamefont {T.~W.}\ \bibnamefont {{Watts}}}, \bibinfo {author} {\bibfnamefont {Y.}~\bibnamefont {{Chen}}}, \bibinfo {author} {\bibfnamefont {A.}~\bibnamefont {{Sarma}}},\ and\ \bibinfo {author} {\bibfnamefont {P.~L.}\ \bibnamefont {{McMahon}}},\ }\bibfield  {title} {\bibinfo {title} {{Linear-depth quantum circuits for loading Fourier approximations of arbitrary functions}},\ }\href {https://doi.org/10.48550/arXiv.2302.03888} {\bibfield  {journal} {\bibinfo  {journal} {arXiv e-prints}\ ,\ \bibinfo {eid} {arXiv:2302.03888}} (\bibinfo {year} {2023})},\ \Eprint {https://arxiv.org/abs/2302.03888} {arXiv:2302.03888 [quant-ph]} \BibitemShut {NoStop}%
\bibitem [{\citenamefont {Garc\'{\i}a-Molina}\ \emph {et~al.}(2022)\citenamefont {Garc\'{\i}a-Molina}, \citenamefont {Rodr\'{\i}guez-Mediavilla},\ and\ \citenamefont {Garc\'{\i}a-Ripoll}}]{Molina2022}%
  \BibitemOpen
  \bibfield  {author} {\bibinfo {author} {\bibfnamefont {P.}~\bibnamefont {Garc\'{\i}a-Molina}}, \bibinfo {author} {\bibfnamefont {J.}~\bibnamefont {Rodr\'{\i}guez-Mediavilla}},\ and\ \bibinfo {author} {\bibfnamefont {J.~J.}\ \bibnamefont {Garc\'{\i}a-Ripoll}},\ }\bibfield  {title} {\bibinfo {title} {Quantum fourier analysis for multivariate functions and applications to a class of schr\"odinger-type partial differential equations},\ }\href {https://doi.org/10.1103/PhysRevA.105.012433} {\bibfield  {journal} {\bibinfo  {journal} {Phys. Rev. A}\ }\textbf {\bibinfo {volume} {105}},\ \bibinfo {pages} {012433} (\bibinfo {year} {2022})}\BibitemShut {NoStop}%
\bibitem [{\citenamefont {{Mottonen}}\ \emph {et~al.}(2004)\citenamefont {{Mottonen}}, \citenamefont {{Vartiainen}}, \citenamefont {{Bergholm}},\ and\ \citenamefont {{Salomaa}}}]{Mottonen2004}%
  \BibitemOpen
  \bibfield  {author} {\bibinfo {author} {\bibfnamefont {M.}~\bibnamefont {{Mottonen}}}, \bibinfo {author} {\bibfnamefont {J.~J.}\ \bibnamefont {{Vartiainen}}}, \bibinfo {author} {\bibfnamefont {V.}~\bibnamefont {{Bergholm}}},\ and\ \bibinfo {author} {\bibfnamefont {M.~M.}\ \bibnamefont {{Salomaa}}},\ }\bibfield  {title} {\bibinfo {title} {{Transformation of quantum states using uniformly controlled rotations}},\ }\href@noop {} {\bibfield  {journal} {\bibinfo  {journal} {arXiv e-prints}\ ,\ \bibinfo {eid} {quant-ph/0407010}} (\bibinfo {year} {2004})},\ \Eprint {https://arxiv.org/abs/quant-ph/0407010} {arXiv:quant-ph/0407010 [quant-ph]} \BibitemShut {NoStop}%
\bibitem [{\citenamefont {Kiani}\ \emph {et~al.}(2022)\citenamefont {Kiani}, \citenamefont {De~Palma}, \citenamefont {Englund}, \citenamefont {Kaminsky}, \citenamefont {Marvian},\ and\ \citenamefont {Lloyd}}]{LLoyd2022}%
  \BibitemOpen
  \bibfield  {author} {\bibinfo {author} {\bibfnamefont {B.~T.}\ \bibnamefont {Kiani}}, \bibinfo {author} {\bibfnamefont {G.}~\bibnamefont {De~Palma}}, \bibinfo {author} {\bibfnamefont {D.}~\bibnamefont {Englund}}, \bibinfo {author} {\bibfnamefont {W.}~\bibnamefont {Kaminsky}}, \bibinfo {author} {\bibfnamefont {M.}~\bibnamefont {Marvian}},\ and\ \bibinfo {author} {\bibfnamefont {S.}~\bibnamefont {Lloyd}},\ }\bibfield  {title} {\bibinfo {title} {Quantum advantage for differential equation analysis},\ }\href {https://doi.org/10.1103/PhysRevA.105.022415} {\bibfield  {journal} {\bibinfo  {journal} {Phys. Rev. A}\ }\textbf {\bibinfo {volume} {105}},\ \bibinfo {pages} {022415} (\bibinfo {year} {2022})}\BibitemShut {NoStop}%
\bibitem [{\citenamefont {{Nakaji}}\ and\ \citenamefont {{Yamamoto}}(2021)}]{ALA}%
  \BibitemOpen
  \bibfield  {author} {\bibinfo {author} {\bibfnamefont {K.}~\bibnamefont {{Nakaji}}}\ and\ \bibinfo {author} {\bibfnamefont {N.}~\bibnamefont {{Yamamoto}}},\ }\bibfield  {title} {\bibinfo {title} {{Expressibility of the alternating layered ansatz for quantum computation}},\ }\href {https://doi.org/10.22331/q-2021-04-19-434} {\bibfield  {journal} {\bibinfo  {journal} {Quantum}\ }\textbf {\bibinfo {volume} {5}},\ \bibinfo {pages} {434} (\bibinfo {year} {2021})},\ \Eprint {https://arxiv.org/abs/2005.12537} {arXiv:2005.12537 [quant-ph]} \BibitemShut {NoStop}%
\bibitem [{\citenamefont {Øksendal}(1998)}]{QuantFinance}%
  \BibitemOpen
  \bibfield  {author} {\bibinfo {author} {\bibfnamefont {B.}~\bibnamefont {Øksendal}},\ }\href@noop {} {\emph {\bibinfo {title} {Stochastic Differential Equations : An Introduction with Applications (5th ed.)}}}\ (\bibinfo  {publisher} {Springer},\ \bibinfo {address} {Berlin},\ \bibinfo {year} {1998})\ p.\ \bibinfo {pages} {266–283}\BibitemShut {NoStop}%
\bibitem [{\citenamefont {Ankudinova}\ and\ \citenamefont {Ehrhardt}(2008)}]{Ankudinova2008}%
  \BibitemOpen
  \bibfield  {author} {\bibinfo {author} {\bibfnamefont {J.}~\bibnamefont {Ankudinova}}\ and\ \bibinfo {author} {\bibfnamefont {M.}~\bibnamefont {Ehrhardt}},\ }\bibfield  {title} {\bibinfo {title} {On the numerical solution of nonlinear black–scholes equations},\ }\href {https://doi.org/https://doi.org/10.1016/j.camwa.2008.02.005} {\bibfield  {journal} {\bibinfo  {journal} {Computers \& Mathematics with Applications}\ }\textbf {\bibinfo {volume} {56}},\ \bibinfo {pages} {799} (\bibinfo {year} {2008})},\ \bibinfo {note} {mathematical Models in Life Sciences \& Engineering}\BibitemShut {NoStop}%
\bibitem [{\citenamefont {{Chatterjee}}(2021)}]{Chatterjee2021}%
  \BibitemOpen
  \bibfield  {author} {\bibinfo {author} {\bibfnamefont {S.}~\bibnamefont {{Chatterjee}}},\ }\bibfield  {title} {\bibinfo {title} {{Universality of deterministic KPZ}},\ }\href@noop {} {\bibfield  {journal} {\bibinfo  {journal} {arXiv e-prints}\ ,\ \bibinfo {eid} {arXiv:2102.13131}} (\bibinfo {year} {2021})},\ \Eprint {https://arxiv.org/abs/2102.13131} {arXiv:2102.13131 [math.PR]} \BibitemShut {NoStop}%
\bibitem [{\citenamefont {{Liu}}\ \emph {et~al.}(2022)\citenamefont {{Liu}}, \citenamefont {{Chen}}, \citenamefont {{Shu}}, \citenamefont {{Chye Chew}},\ and\ \citenamefont {{Cheong Khoo}}}]{Yangyang2022}%
  \BibitemOpen
  \bibfield  {author} {\bibinfo {author} {\bibfnamefont {Y.}~\bibnamefont {{Liu}}}, \bibinfo {author} {\bibfnamefont {Z.}~\bibnamefont {{Chen}}}, \bibinfo {author} {\bibfnamefont {C.}~\bibnamefont {{Shu}}}, \bibinfo {author} {\bibfnamefont {S.}~\bibnamefont {{Chye Chew}}},\ and\ \bibinfo {author} {\bibfnamefont {B.}~\bibnamefont {{Cheong Khoo}}},\ }\bibfield  {title} {\bibinfo {title} {{Application of a variational hybrid quantum-classical algorithm to heat conduction equation}},\ }\href@noop {} {\bibfield  {journal} {\bibinfo  {journal} {arXiv e-prints}\ ,\ \bibinfo {eid} {arXiv:2207.14630}} (\bibinfo {year} {2022})},\ \Eprint {https://arxiv.org/abs/2207.14630} {arXiv:2207.14630 [quant-ph]} \BibitemShut {NoStop}%
\bibitem [{\citenamefont {{Rosenberg}}\ \emph {et~al.}(2022)\citenamefont {{Rosenberg}}, \citenamefont {{Ginsparg}},\ and\ \citenamefont {{McMahon}}}]{Eliott}%
  \BibitemOpen
  \bibfield  {author} {\bibinfo {author} {\bibfnamefont {E.}~\bibnamefont {{Rosenberg}}}, \bibinfo {author} {\bibfnamefont {P.}~\bibnamefont {{Ginsparg}}},\ and\ \bibinfo {author} {\bibfnamefont {P.~L.}\ \bibnamefont {{McMahon}}},\ }\bibfield  {title} {\bibinfo {title} {{Experimental error mitigation using linear rescaling for variational quantum eigensolving with up to 20 qubits}},\ }\href {https://doi.org/10.1088/2058-9565/ac3b37} {\bibfield  {journal} {\bibinfo  {journal} {Quantum Science and Technology}\ }\textbf {\bibinfo {volume} {7}},\ \bibinfo {eid} {015024} (\bibinfo {year} {2022})},\ \Eprint {https://arxiv.org/abs/2106.01264} {arXiv:2106.01264 [quant-ph]} \BibitemShut {NoStop}%
\bibitem [{\citenamefont {{Guseynov}}\ \emph {et~al.}(2022)\citenamefont {{Guseynov}}, \citenamefont {{Zhukov}}, \citenamefont {{Pogosov}},\ and\ \citenamefont {{Lebedev}}}]{Guseynov2022}%
  \BibitemOpen
  \bibfield  {author} {\bibinfo {author} {\bibfnamefont {N.~M.}\ \bibnamefont {{Guseynov}}}, \bibinfo {author} {\bibfnamefont {A.~A.}\ \bibnamefont {{Zhukov}}}, \bibinfo {author} {\bibfnamefont {W.~V.}\ \bibnamefont {{Pogosov}}},\ and\ \bibinfo {author} {\bibfnamefont {A.~V.}\ \bibnamefont {{Lebedev}}},\ }\bibfield  {title} {\bibinfo {title} {{Depths analysis of variational quantum algorithms for heat equation}},\ }\href {https://doi.org/10.48550/arXiv.2212.12375} {\bibfield  {journal} {\bibinfo  {journal} {arXiv e-prints}\ ,\ \bibinfo {eid} {arXiv:2212.12375}} (\bibinfo {year} {2022})},\ \Eprint {https://arxiv.org/abs/2212.12375} {arXiv:2212.12375 [quant-ph]} \BibitemShut {NoStop}%
\bibitem [{\citenamefont {{Pool}}\ \emph {et~al.}(2024)\citenamefont {{Pool}}, \citenamefont {{Somoza}}, \citenamefont {{Mc Keever}}, \citenamefont {{Lubasch}},\ and\ \citenamefont {{Horstmann}}}]{lubasch_full_evo}%
  \BibitemOpen
  \bibfield  {author} {\bibinfo {author} {\bibfnamefont {A.~J.}\ \bibnamefont {{Pool}}}, \bibinfo {author} {\bibfnamefont {A.~D.}\ \bibnamefont {{Somoza}}}, \bibinfo {author} {\bibfnamefont {C.}~\bibnamefont {{Mc Keever}}}, \bibinfo {author} {\bibfnamefont {M.}~\bibnamefont {{Lubasch}}},\ and\ \bibinfo {author} {\bibfnamefont {B.}~\bibnamefont {{Horstmann}}},\ }\bibfield  {title} {\bibinfo {title} {{Nonlinear dynamics as a ground-state solution on quantum computers}},\ }\href {https://doi.org/10.48550/arXiv.2403.16791} {\bibfield  {journal} {\bibinfo  {journal} {arXiv e-prints}\ ,\ \bibinfo {eid} {arXiv:2403.16791}} (\bibinfo {year} {2024})},\ \Eprint {https://arxiv.org/abs/2403.16791} {arXiv:2403.16791 [quant-ph]} \BibitemShut {NoStop}%
\end{thebibliography}%

\newpage

\appendix

\section{Real-Valued ZGR-QFT Ansatz}
\label{appendix:ZGRQFT}

To construct the real-valued ZGR-QFT, we wish to parameterize a truncated real-valued Fourier series. Consider the following partial Fourier series:
\begin{align}
    f(x) \, = \, \sum_{k=0}^{M-1} \, a_{k} \cos(2\pi k x) \, + \, \sum_{k=1}^{M-1} \, b_{k} \sin(2\pi k x) \, ,
\end{align}
where $M = 2^{m}$. Note that we can equivalently write the above function as 
\begin{align}
    f(x) \, = \, \sum_{k=0}^{M-1} \, c_{k} e^{-i2\pi k x} \, + \, \sum_{k=1}^{M-1} \, c^{*}_{k} e^{i2\pi k x} \, ,\label{eq-fourier-c}
\end{align}
where $c_{0} \, = \, a_{0}\, $ and $c_{k} \, = \, \frac{1}{2}\left(a_{k} + i b_{k}\right) \, $ for $k \ne 0$.

In accordance with the real-valued discrete Fourier transform, the $n$-qubit state $\ket{f}$ can be written as 
\begin{align}
    \ket{f} \, = \, \sqrt{N} \,\, \qft^{-1} \ket{\psi_{f}} \, ,\label{eq-real-ans}
\end{align}
where $N = 2^{n}$ and the (unnormalized) state $\ket{\psi_{f}}$ is given by
\begin{align}
    \ket{\psi_{f}} \, = \, \sum_{k=0}^{M-1} \, c_{k} \, \ket{k} \, + \, \sum_{k=1}^{M-1} \, c^{*}_{k} \, \ket{N-k} \, .  \label{eq-psif}
\end{align}

\subsection{Quantum circuit to prepare $\ket{\psi_{f}}$}

Consider the following $m$ qubits quantum states which encode the Fourier coefficients:
\begin{align}
    \ket{c} \, =\, \, \frac{1}{\mathcal{N}_{c}} \, \left[ \frac{c_{0}}{\sqrt{2}} \, \ket{0} \, + \, \sum_{k=1}^{M-1} \, c_{k} \, \ket{k} \right] \, ,\label{eq-state-c}\\
    \ket{c^{*}} \, =\, \, \frac{1}{\mathcal{N}_{c}} \, \left[ \frac{c_{0}}{\sqrt{2}} \, \ket{0} \, + \, \sum_{k=1}^{M-1} \, c^{*}_{k} \, \ket{k} \right] \, , \label{eq-state-cc}
\end{align}
where the normalization constant is
\begin{equation}
    \left(\mathcal{N}_{c}\right)^{2} \, = \, \frac{c^{2}_{0}}{2} \, + \, \sum_{k=1}^{M-1} |c_{k}|^{2} \, .
\end{equation}
The states $\ket{c}$ and $\ket{c^{*}}$ can efficiently be prepared using the ZGR unitary, which we denote as $\hat{U}_{c}(\vec{\theta}_{y}, \vec{\theta}_{z}, \alpha)$ (shown in Fig. 1b of \cite{Watts2023}). These ZGR unitaries consist of a cascade of uniformly controlled rotations, as described in \cite{Mottonen2004}. Here, $\vec{\theta}_{y/z}$ contains all of the rotation parameters for the $\hat{R}_{y/z}$ gates, while $\alpha$ is a global phase. It is straightforward to see from the formulae of ZGR parameters in \cite{Mottonen2004} that the ZGR parameters to prepare the conjugate state $\ket{c^{*}}$ are the same as the parameters for $\ket{c}$, but with $\vec{\theta}_{z}$ and $\alpha$ negated:
\begin{align}
    &\ket{c} \, = \, \hat{U}_{c}(\vec{\theta}_{y}, \vec{\theta}_{z}, \alpha) \ket{0}^{\otimes m} \, \\ \, &\ket{c^{*}} \, = \, \hat{U}_{c}(\vec{\theta}_{y}, -\vec{\theta}_{z}, -\alpha) \ket{0}^{\otimes m} \, .
\end{align}

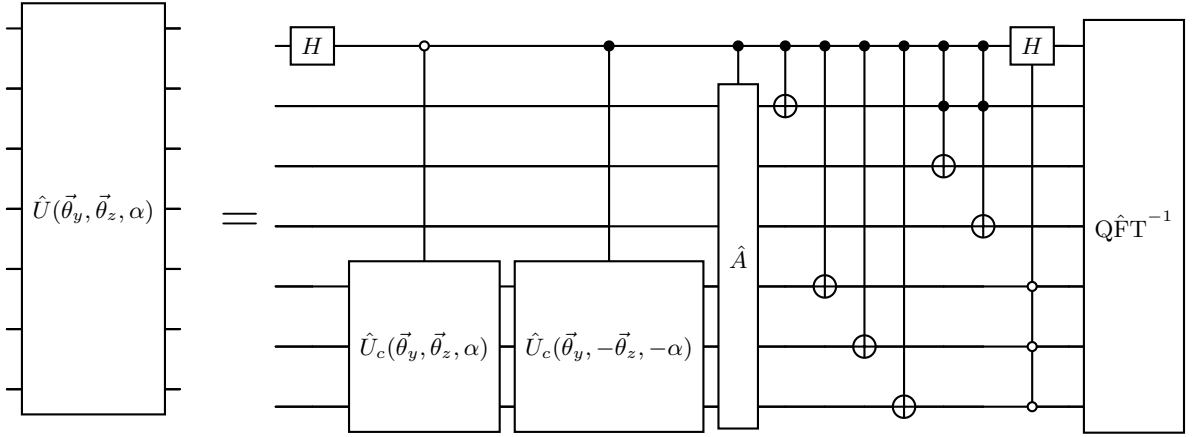
\begin{figure}
\begin{tikzpicture}
\node at (0, 0)[scale=1]{
\begin{quantikz}[column sep=0.2cm, row sep={0.8cm,between origins}]
    \qw & \gate[wires=7]{\hat{U}(\vec{\theta}_{y}, \vec{\theta}_{z}, \alpha)} & \qw\\
    \qw & \qw & \qw\\
    \qw & \qw & \qw\\
    \qw & \qw & \qw\\
    \qw & \qw & \qw\\
    \qw & \qw & \qw\\
    \qw & \qw & \qw\\
\end{quantikz} 
\hspace{0.3cm}
{\huge =}
\begin{quantikz}[column sep=0.2cm, row sep={0.8cm,between origins}]
\qw & \gate{H} & \octrl{4} & \ctrl{4} & \ctrl{1} & \ctrl{1} & \ctrl{4} & \ctrl{5} & \ctrl{6} & \ctrl{1} & \ctrl{1} & \gate{H} & \qw & \gate[wires=7]{\qft^{-1}}\\
     \qw & \qw & \qw & \qw & \gate[wires=6]{\hat{A}} & \targ & \qw & \qw & \qw & \qw & \ctrl{1} & \ctrl{2} & \qw & \qw & \qw\\
     \qw & \qw & \qw & \qw & \linethrough & \qw & \qw & \qw & \qw & \targ & \qw & \qw & \qw & \qw & \qw \\
     \qw & \qw & \qw & \qw & \linethrough & \qw & \qw & \qw & \qw & \qw & \targ & \qw & \qw & \qw & \qw\\
     \qw & \qw & \gate[wires=3]{\hat{U}_{c}(\vec{\theta}_{y}, \vec{\theta}_{z}, \alpha)} & \gate[wires=3]{\hat{U}_{c}(\vec{\theta}_{y}, -\vec{\theta}_{z}, -\alpha)} & \qw & \qw & \targ & \qw & \qw & \qw & \qw & \qw & \octrl{-4} & \qw & \qw\\
     \qw & \qw & \qw & \qw & \qw & \qw & \qw & \targ & \qw & \qw & \qw & \qw & \octrl{-1} & \qw & \qw \\
     \qw & \qw & \qw & \qw & \qw & \qw & \qw & \qw & \targ & \qw & \qw & \qw & \octrl{-1} & \qw & \qw 
\end{quantikz}
};

\end{tikzpicture}
\caption{Real-Valued ZGR-QFT Ansatz Circuit Diagram. Displayed here with $n=7$, $m=3$. The generalization for arbitrary $m$ and $n\geq m+2$ is straightforward. Note that the lines through the Adder gate means that the Adder is not acting on the corresponding qubits. One can straightforwardly verify that the gates before the QFT prepare a state proportional to $\ket{\psi_f}$.}
\label{fig:zgrqft}
\end{figure}

Given the unitary $\hat{U}_{c}$, the circuit to produce a state proportional to $\ket{f}$ is given in Fig. \ref{fig:zgrqft}. To convert this into a variational ansatz, we simply promote the variables $(\vec{\theta}_{y}, \vec{\theta}_{z}, \alpha)$ to variational parameters.

\onecolumngrid

\twocolumngrid

\subsection{Comments on the Adder operator}

The quantum circuit that we have proposed contains the $(m+1)$-qubit Adder operator. One way to implement this Adder operator is using $O(m)$ CNOTs and Toffoli gates but this implementation requires $m-1$ ancilla qubits. Another way to implement this is using the phase shift gates and $(m+1)$-qubit QFT and its inverse. This method does not require ancilla qubits but require $O(m^{2})$ CNOT gates. 

In our code, we make use of both of these implementations. If $n>2m$, then we use the implementation of Adder with ancilla qubits and use the remaining $(n-2-m)$ qubits as ancilla. On the other hand, if $n\le 2m$, we implement the Adder operator using the phase shift and the Adder operator.  

\section{Universal Layered Ansatz}
\label{appendix:ULA}

\begin{figure}[h]
\begin{tikzpicture}

\node at (5, 0)[scale=1]{
\begin{quantikz}[column sep=0.15cm, row sep={0.8cm,between origins}]
\qw & \gate[wires=5]{\hat{U}(\vec{\theta})} & \qw\\
\qw & \qw & \qw\\
\qw & \qw & \qw\\
\qw & \qw & \qw\\
\qw & \qw & \qw\\
\end{quantikz} 
\hspace{0.3cm}
{\huge =}
\begin{quantikz}[column sep=0.15cm, row sep={0.8cm,between origins}]
\qw & \gate[wires=2]{SO(4)} & \qw & \gate[wires=2]{SO(4)} & \qw & \qw\\
\qw & \qw & \gate[wires=2]{SO(4)} & \qw & \gate[wires=2]{SO(4)} & \qw\\
\qw & \gate[wires=2]{SO(4)} & \qw & \gate[wires=2]{SO(4)} & \qw & \qw\\
\qw & \qw & \gate[wires=2]{SO(4)} & \qw & \gate[wires=2]{SO(4)} & \qw\\
\qw & \qw & \qw & \qw & \qw & \qw\\
\end{quantikz};
};

\node at (5, 5)[scale=1] {
\begin{quantikz}[column sep=0.15cm, row sep={0.8cm,between origins}]
\qw & \gate[wires=2]{SO(4)} & \qw \\
\qw & \qw & \qw \\
\end{quantikz}
\hspace{0.3cm}
{\huge =}
\begin{quantikz}[column sep=0.15cm, row sep={0.8cm,between origins}]
\qw & \gate{R_y(\theta_1)} & \ctrl{1} & \gate{R_y(\theta_3)} & \ctrl{1} & \gate{R_y(\theta_5)} & \qw \\
\qw & \gate{R_y(\theta_2)} & \targ{} & \gate{R_y(\theta_4)} & \targ{} & \gate{R_y(\theta_6)} & \qw\\
\end{quantikz}
};

\end{tikzpicture};
\caption{ULA Circuit Diagram. Here the ULA with $n=5$ qubits and depth $d=2$ is displayed. Each SO(4) gate contains six different parameters, giving a total of $6(n-1)d$ parameters.}
\end{figure}
\FloatBarrier

\section{1D Linear and Nonlinear Black--Scholes Equation Cost-Functions}
\label{appendix:1DBSE}

We will henceforth denote the parameters of $\ket{\chi}$ by $\theta_0$ and $\boldsymbol{\theta}$, and the parameters of $\ket{V}$ by $\lambda_0$ and $\boldsymbol{\lambda}$, such that $\ket{\chi}=\theta_0\hat{U}(\boldsymbol{\theta})\ket{0}=\theta_0\ket{\phi}$ and $\ket{V}=\lambda_0\hat{U}(\boldsymbol{\lambda})\ket{0}=\lambda_0\ket{\psi}$, where we have defined $\ket{\phi}=\hat{U}(\boldsymbol{\theta})\ket{0}$ and $\ket{\psi}=\hat{U}(\boldsymbol{\lambda})\ket{0}$, and we have dropped the explicit dependence on the rotational parameters. Once again denoting the parameters of the previous timestep with a tilde, the cost-function for $\ket{\chi}$ is as follows:

\begin{equation}
\begin{gathered}
    C_\chi = ||\ket{\chi}-(\frac{\hat{\partial}^2}{\partial x^2} - \frac{\hat{\partial}}{\partial x})\ket{\tilde{V}}||^2 \\
    = \braket{\chi - (\frac{\hat{\partial}^2}{\partial x^2} - \frac{\hat{\partial}}{\partial x})\tilde{V} | \chi - (\frac{\hat{\partial}^2}{\partial x^2} - \frac{\hat{\partial}}{\partial x})\tilde{V}}
\end{gathered}
\end{equation}

\noindent where we are once again using the naive distance measure between $\ket{\chi}$ and the desired state. After making the substitutions $\frac{\hat{\partial}^2}{\partial x^2}=\frac{4^{n}}{L^2}(\hat{A}+\hat{A}^\dag - 2\mathds{1})$, $\frac{\hat{\partial}}{\partial x} = \frac{2^{n-1}}{L}(\hat{A}-\hat{A}^\dag)$, $\ket{\chi}=\theta_0\ket{\phi}$, and $\ket{\tilde{V}}=\tilde{\lambda}_0\ket{\tilde{\psi}}$, and expanding and simplifying, we get

\begin{equation}
\begin{gathered}
    C_\chi = \theta_0^2 + 2\theta_0\tilde{\lambda}_0\operatorname{Re}\{(\frac{2^{n-1}}{L}-\frac{4^n}{L^2})\braket{\phi|\hat{A}|\tilde{\psi}} \\-  (\frac{2^{n-1}}{L}+\frac{4^n}{L^2})\braket{\phi|\hat{A}^\dag|\tilde{\psi}} + 2\cdot\frac{4^n}{L^2}\braket{\phi|\tilde{\psi}}\}+const.
\end{gathered}
\end{equation}

\noindent The three expectations in $C_\chi$ can be calculated efficiently using the Hadamard test. Now that we have the state $\ket{\chi}$, we can use the Diagonal operator, as described in Ref.~\cite{lubasch}, to multiply by the nonlinear volatility. Using this operator, we can generate the "quantum" version of the differential operator $\hat{O}$.

\begin{equation}
\begin{gathered}
\hat{O}=r + (\frac{\sigma^2}{2}-r)\frac{\partial}{\partial x}+\frac{\sigma^2}{2}\frac{\partial^2}{\partial x^2} \\
= r\mathds{1} + (\frac{\sigma_0^2}{2}(1+e^{r(T-t)}a^2\hat{D}_\chi)-r)\frac{\hat{\partial}}{\partial x} \\- \frac{\sigma_0^2}{2}(1+e^{r(T-t)}a^2\hat{D}_\chi)\frac{\hat{\partial}^2}{\partial x^2} 
\end{gathered}
\end{equation}

\noindent Substituting into Eq.~\ref{costfunc}, the cost-function for $\ket{V}$ takes the form

\begin{equation}
\begin{gathered}
    C_V = \langle ((1-r\tau)\mathds{1}-(\frac{1}{2}(\sigma_0^2[1+e^{r(T-t)}a^2\hat{D}_\chi])-r)\tau\frac{\hat{\partial}}{\partial x} \\ +\frac{1}{2}(\sigma_0^2[1+e^{r(T-t)}a^2\hat{D}_\chi])\tau\frac{\hat{\partial}^2}{\partial x^2})V - \tilde{V} | \\
    ((1-r\tau)\mathds{1}-(\frac{1}{2}(\sigma_0^2[1+e^{r(T-t)}a^2\hat{D}_\chi])-r)\tau\frac{\hat{\partial}}{\partial x} \\ +\frac{1}{2}(\sigma_0^2[1+e^{r(T-t)}a^2\hat{D}_\chi])\tau\frac{\hat{\partial}^2}{\partial x^2})V - \tilde{V} \rangle
\end{gathered}
\end{equation}

\noindent Noting that $\hat{D}_\chi=\theta_0\hat{D}_\phi$, we get 

\begin{equation}
\begin{gathered}
    C_V = \langle \left[\alpha\mathds{1}-\beta\frac{\hat{\partial}}{\partial x}+\gamma\frac{\hat{\partial}^2}{\partial x^2}-\epsilon\hat{D}_\phi\frac{\hat{\partial}}{\partial x}+\epsilon\hat{D}_\phi\frac{\hat{\partial}^2}{\partial x^2}\right]V - \tilde{V} | \\
    \left[\alpha\mathds{1}-\beta\frac{\hat{\partial}}{\partial x}+\gamma\frac{\hat{\partial}^2}{\partial x^2}-\epsilon\hat{D}_\phi\frac{\hat{\partial}}{\partial x}+\epsilon\hat{D}_\phi\frac{\hat{\partial}^2}{\partial x^2}\right]V - \tilde{V} \rangle
\end{gathered}
\end{equation}

\noindent where we have defined the constants $\alpha=1-r\tau$, $\beta=(\frac{1}{2}\sigma_0^2-r)\tau$, $\gamma=\frac{1}{2}\sigma_0^2\tau$, and $\epsilon=\frac{1}{2}\sigma_0^2e^{r(T-t)}a^2\theta_0\tau$. This cost-function can be factored as such:

\onecolumngrid

\begin{equation}
\begin{gathered}
    C_V = \langle (\alpha\mathds{1}-\beta\frac{\hat{\partial}}{\partial x}+\gamma\frac{\hat{\partial}^2}{\partial x^2})V - \tilde{V}| (\alpha\mathds{1}-\beta\frac{\hat{\partial}}{\partial x}+\gamma\frac{\hat{\partial}^2}{\partial x^2})V - \tilde{V} \rangle\\
    + 2\operatorname{Re}\{\braket{(\alpha\mathds{1}-\beta\frac{\hat{\partial}}{\partial x}+\gamma\frac{\hat{\partial}^2}{\partial x^2})V | (-\epsilon\hat{D}_\phi\frac{\hat{\partial}}{\partial x}+\epsilon\hat{D}_\phi\frac{\hat{\partial}^2}{\partial x^2})V}\}\\
    + \braket{(-\epsilon\hat{D}_\phi\frac{\hat{\partial}}{\partial x}+\epsilon\hat{D}_\phi\frac{\hat{\partial}^2}{\partial x^2})V|(-\epsilon\hat{D}_\phi\frac{\hat{\partial}}{\partial x}+\epsilon\hat{D}_\phi\frac{\hat{\partial}^2}{\partial x^2})V} \\
    - 2\operatorname{Re}\{\braket{(-\epsilon\hat{D}_\phi\frac{\hat{\partial}}{\partial x}+\epsilon\hat{D}_\phi\frac{\hat{\partial}^2}{\partial x^2})V | \tilde{V}}\}
\end{gathered}
\end{equation}

\noindent Note that the first term is identically equal to the linear Backwards Euler Black--Scholes cost-function. All three terms can be simplified by substituting the operator identities for the derivatives, and expanding the inner product. In the end, we get:

\begin{equation}
\begin{gathered}
    C_{linear} = \tilde{\lambda}_0^2+\lambda_0^2(\alpha^2+\frac{2\beta^2 4^{n-1}}{L^2} + \frac{6\gamma^2 16^n}{L^4} - \frac{4\alpha\gamma 4^n}{L^2}) \\
    + 2\lambda_0\operatorname{Re}\{\lambda_0(\frac{\gamma^2 16^n}{L^4} - \frac{\beta^2 4^{n-1}}{L^2})\braket{\psi|\hat{A}^2|\psi}
    + \lambda_0(\frac{2\alpha\gamma 4^n}{L^2} - \frac{4\gamma^2 16^n}{L^4})\braket{\psi|\hat{A}|\psi} \\
    - \tilde{\lambda}_0(\frac{\beta 2^{n-1}}{L} + \frac{\gamma 4^n}{L^2}) \braket{\psi|\hat{A}|\tilde{\psi}} 
    + \tilde{\lambda}_0(\frac{\beta 2^{n-1}}{L} - \frac{\gamma 4^n}{L^2}) \braket{\psi|\hat{A}^\dag|\tilde{\psi}}
    + \tilde{\lambda}_0(\frac{2\gamma 4^n}{L^2} - \alpha)\braket{\psi|\tilde{\psi}}\}
\end{gathered}
\end{equation}

\begin{equation}
\begin{gathered}
        C_{2} = 2\lambda_0\operatorname{Re}\{\lambda_0(\frac{\alpha\epsilon4^n}{L^2}-\frac{\alpha\epsilon2^{n-1}}{L} +\frac{2\gamma\epsilon2^{n-1}4^n}{L^3}-\frac{2\gamma\epsilon16^n}{L^4})\braket{\psi|\hat{D}_\phi\hat{A}|\psi} \\ 
    + \lambda_0(\frac{\alpha\epsilon4^n}{L^2}+\frac{\alpha\epsilon2^{n-1}}{L} -\frac{2\gamma\epsilon2^{n-1}4^n}{L^3}-\frac{2\gamma\epsilon16^n}{L^4})\braket{\psi|\hat{D}_\phi\hat{A}^\dag|\psi} \\ 
    + \lambda_0(\frac{4\gamma\epsilon16^n}{L^4}-\frac{2\alpha\epsilon4^n}{L^2})\braket{\psi|\hat{D}_\phi|\psi} \\
    + \lambda_0(\frac{\beta\epsilon4^{n-1}}{L^2}-\frac{\beta\epsilon2^{n-1}4^n}{L^3} +\frac{\gamma\epsilon16^n}{L^4}-\frac{\gamma\epsilon2^{n-1}4^n}{L^3})\braket{\psi|\hat{A}^\dag\hat{D}_\phi\hat{A}|\psi} \\
    + \lambda_0(-\frac{\beta\epsilon4^{n-1}}{L^2}-\frac{\beta\epsilon2^{n-1}4^n}{L^3} +\frac{\gamma\epsilon16^n}{L^4}+\frac{\gamma\epsilon2^{n-1}4^n}{L^3})\braket{\psi|\hat{A}^\dag\hat{D}_\phi\hat{A}^\dag|\psi} \\
    + \lambda_0(-\frac{\beta\epsilon4^{n-1}}{L^2}+\frac{\beta\epsilon2^{n-1}4^n}{L^3} +\frac{\gamma\epsilon16^n}{L^4}-\frac{\gamma\epsilon2^{n-1}4^n}{L^3})\braket{\psi|\hat{A}\hat{D}_\phi\hat{A}|\psi} \\
    + \lambda_0(\frac{\beta\epsilon4^{n-1}}{L^2}+\frac{\beta\epsilon2^{n-1}4^n}{L^3} +\frac{\gamma\epsilon16^n}{L^4}+\frac{\gamma\epsilon2^{n-1}4^n}{L^3})\braket{\psi|\hat{A}\hat{D}_\phi\hat{A}^\dag|\psi} \\
    + \lambda_0(\frac{2\beta\epsilon2^{n-1}4^n}{L^3}-\frac{2\gamma\epsilon16^n}{L^4})\braket{\psi|\hat{A}^\dag\hat{D}_\phi|\psi}
    + \lambda_0(-\frac{2\beta\epsilon2^{n-1}4^n}{L^3}-\frac{2\gamma\epsilon16^n)}{L^4}\braket{\psi|\hat{A}\hat{D}_\phi|\psi} \}
\end{gathered}
\end{equation}

\begin{equation}
    \begin{gathered}
        C_{3} = 2\lambda_0\operatorname{Re}\{\lambda_0 (\frac{-\epsilon^2 4^{n-1}}{L^2} + \frac{\epsilon^2 16^n}{L^4})\braket{\psi | \hat{A} \hat{D}_\phi^\dag \hat{D}_\phi \hat{A} | \psi} \\
        + \lambda_0(\frac{\epsilon^2 4^{n-1}}{2L^2} + \frac{\epsilon^2 4^n 2^{n-1}}{L^3} + \frac{\epsilon^2 16^n}{2L^4})\braket{\psi | \hat{A} \hat{D}_\phi^\dag \hat{D}_\phi \hat{A}^\dag | \psi} \\
        + \lambda_0(\frac{\epsilon^2 4^{n-1}}{2L^2} - \frac{\epsilon^2 4^n 2^{n-1}}{L^3} + \frac{\epsilon^2 16^n}{2L^4})\braket{\psi | \hat{A}^\dag \hat{D}_\phi^\dag \hat{D}_\phi \hat{A} | \psi} \\
        + \lambda_0(\frac{2\epsilon^2 4^n 2^{n-1}}{L^3} - \frac{2\epsilon^2 16^n}{L^4})\braket{\psi | \hat{D}_\phi^\dag \hat{D}_\phi \hat{A} | \psi}
        + \lambda_0(\frac{-2\epsilon^2 4^n 2^{n-1}}{L^3} - \frac{2\epsilon^2 16^n}{L^4})\braket{\psi | \hat{D}_\phi^\dag \hat{D}_\phi \hat{A}^\dag | \psi} \\
        + \lambda_0(\frac{2\epsilon^2 16^n}{L^4})\braket{\psi | \hat{D}_\phi^\dag \hat{D}_\phi | \psi} \}
    \end{gathered}
\end{equation}

\begin{equation}
    \begin{gathered}
        C_{4} = 2\lambda_0\operatorname{Re}\{\tilde{\lambda}_0 (\frac{\epsilon^2 2^{n-1}}{L} - \frac{\epsilon^2 4^n}{L^2})\braket{\tilde{\psi} | \hat{D}_\phi \hat{A} | \psi} \\
        + \tilde{\lambda}_0(-\frac{\epsilon^2 2^{n-1}}{L} - \frac{\epsilon^2 4^n}{L^2})\braket{\tilde{\psi} | \hat{D}_\phi \hat{A}^\dag | \psi} \\
        + \tilde{\lambda}_0(\frac{2\epsilon^2 4^{n}}{L^2})\braket{\tilde{\psi} | \hat{D}_\phi| \psi} \}
    \end{gathered}
\end{equation}

\section{2D Linear Black--Scholes Equation Cost-Function}
\label{appendix:2DBSE}

First, defining $\alpha = 1-r\tau$, $\beta_x = (\frac{1}{2}\sigma_x^2-r)\tau$, $\beta_y = (\frac{1}{2}\sigma_y^2-r)\tau$, $\gamma_x = \frac{1}{2}\sigma_x^2\tau$, $\gamma_y = \frac{1}{2}\sigma_y^2\tau$, $\gamma_{xy} = \frac{1}{2}\rho\sigma_x\sigma_y\tau$,

\begin{equation}
\begin{gathered}
    C_V = \langle \left[\alpha\mathds{1} - \beta_x \frac{\hat{\partial}}{\partial x} - \beta_y \frac{\hat{\partial}}{\partial y} + \gamma_x \frac{\hat{\partial}^2}{\partial x^2} + \gamma_y \frac{\hat{\partial}^2}{\partial y^2} + \gamma_{xy} \frac{\hat{\partial}^2}{\partial x \partial y}\right]V - \tilde{V} | \\
    \left[\alpha\mathds{1} - \beta_x \frac{\hat{\partial}}{\partial x} - \beta_y \frac{\hat{\partial}}{\partial y} + \gamma_x \frac{\hat{\partial}^2}{\partial x^2} + \gamma_y \frac{\hat{\partial}^2}{\partial y^2} + \gamma_{xy} \frac{\hat{\partial}^2}{\partial x \partial y}\right]V - \tilde{V} \rangle
\end{gathered}
\end{equation}

\noindent following from Eq.~\ref{costfunc}. After plugging in the definitions of $\ket{V}$, $\ket{\tilde{V}}$, and the derivatives, we are ultimately left with

\begin{equation}
\begin{gathered}
    C_V = \tilde{\lambda}_0^2 + \lambda_0^2(\alpha^2 + \frac{6\gamma_x^2 16^{n_x}}{L_x^4} + \frac{6\gamma_y^2 16^{n_y}}{L_y^4} + \frac{4^{n_x}\beta_x^2}{2L_x^2} + \frac{4^{n_y}\beta_y^2}{2L_y^2} - \frac{4\alpha\gamma_x 4^{n_x}}{L_x^2} - \frac{4\alpha\gamma_y 4^{n_y}}{L_y^2} + \frac{4^{n_x}4^{n_y}\gamma_{xy}^2}{4L_x^2 L_y^2} + \frac{8\gamma_x\gamma_y 4^{n_x}4^{n_y}}{L_x^2 L_y^2}) \\
    + 2\lambda_0\operatorname{Re}\{ 
    \tilde{\lambda}_0(-\alpha + \frac{2 \gamma_x 4^{n_x}}{L_x^2} + \frac{2 \gamma_y 4^{n_y}}{L_y^2})\braket{\psi|\tilde{\psi}}
    +\tilde{\lambda}_0(-\frac{2^{n_x-1}\beta_x}{L_x} - \frac{4^{n_x}\gamma_x}{L_x^2})\braket{\psi|\hat{A}_x|\tilde{\psi}} \\
    +\tilde{\lambda}_0(\frac{2^{n_x-1}\beta_x}{L_x} - \frac{4^{n_x}\gamma_x}{L_x^2})\braket{\psi|\hat{A}^\dag_x|\tilde{\psi}}
    +\tilde{\lambda}_0(-\frac{2^{n_y-1}\beta_y}{L_y} - \frac{4^{n_y}\gamma_y}{L_y^2})\braket{\psi|\hat{A}_y|\tilde{\psi}} \\
    +\tilde{\lambda}_0(\frac{2^{n_y-1}\beta_y}{L_y} - \frac{4^{n_y}\gamma_y}{L_y^2})\braket{\psi|\hat{A}^\dag_y|\tilde{\psi}}
    +\lambda_0(-\frac{4^{n_x}4^{n_y}\gamma_{xy}^2}{8L_x^2 L_y^2} - \frac{4^{n_x}\beta_x^2}{4L_x^2}+\frac{16^{n_x}\gamma_x^2}{L_x^4})\braket{\psi|\hat{A}_x^2|\psi} \\
    +\lambda_0(\frac{8^{n_x}2^{n_y}\gamma_x \gamma_{xy}}{2L_x^3 L_y})\braket{\psi|\hat{A}_x^2 \hat{A}_y|\psi}
    +\lambda_0(\frac{4^{n_x}4^{n_y}\gamma_{xy}^2}{16L_x^2 L_y^2})\braket{\psi|\hat{A}_x^2 \hat{A}_y^2|\psi} \\
    +\lambda_0(-\frac{8^{n_x}2^{n_y}\gamma_x \gamma_{xy}}{2L_x^3 L_y})\braket{\psi|\hat{A}_x^2 \hat{A}^\dag_y|\psi}
    +\lambda_0(\frac{4^{n_x}4^{n_y}\gamma_{xy}^2}{16L_x^2 L_y^2})\braket{\psi|\hat{A}_x^2 (\hat{A}^\dag_y)^2|\psi} \\
    +\lambda_0(-\frac{4^{n_x}4^{n_y}\gamma_{xy}^2}{8L_x^2 L_y^2} - \frac{4^{n_y}\beta_y^2}{4L_y^2}+\frac{16^{n_y}\gamma_y^2}{L_y^4})\braket{\psi|\hat{A}_y^2|\psi} \\
    +\lambda_0(\frac{2^{n_x}8^{n_y}\gamma_y \gamma_{xy}}{2L_x L_y^3})\braket{\psi|\hat{A}_y^2 \hat{A}_x|\psi}
    +\lambda_0(-\frac{2^{n_x}8^{n_y}\gamma_y \gamma_{xy}}{2L_x L_y^3})\braket{\psi|\hat{A}_y^2 \hat{A}^\dag_x|\psi} \\
    +\lambda_0(-\frac{4\cdot 16^{n_x}\gamma_x^2}{L_x^4}-\frac{4\cdot 4^{n_x}4^{n_y}\gamma_x \gamma_y}{L_x^2 L_y^2}+\frac{2\cdot 4^{n_x}\alpha\gamma_x}{L_x^2})\braket{\psi|\hat{A}_x|\psi} \\
    +\lambda_0(-\frac{8^{n_x}2^{n_y}\gamma_x \gamma_{xy}}{L_x^3 L_y}+\frac{2\cdot 4^{n_x}4^{n_y}\gamma_x \gamma_y}{L_x^2 L_y^2}-\frac{2^{n_x}8^{n_y}\gamma_y \gamma_{xy}}{L_x L_y^3}+\frac{2^{n_x}2^{n_y}\alpha\gamma_{xy}}{2L_x L_y}-\frac{2^{n_x}2^{n_y}\beta_x\beta_y}{2L_x L_y})\braket{\psi|\hat{A}_x \hat{A}_y|\psi} \\
    +\lambda_0(\frac{8^{n_x}2^{n_y}\gamma_x \gamma_{xy}}{L_x^3 L_y}+\frac{2\cdot 4^{n_x}4^{n_y}\gamma_x \gamma_y}{L_x^2 L_y^2}+\frac{2^{n_x}8^{n_y}\gamma_y \gamma_{xy}}{L_x L_y^3}-\frac{2^{n_x}2^{n_y}\alpha\gamma_{xy}}{2L_x L_y}+\frac{2^{n_x}2^{n_y}\beta_x\beta_y}{2L_x L_y})\braket{\psi|\hat{A}_x \hat{A}^\dag_y|\psi} \\
    +\lambda_0(-\frac{4\cdot 16^{n_y}\gamma_y^2}{L_y^4}-\frac{4\cdot 4^{n_x}4^{n_y}\gamma_x \gamma_y}{L_x^2 L_y^2}+\frac{2\cdot 4^{n_y}\alpha\gamma_y}{L_y^2})\braket{\psi|\hat{A}_y|\psi}
    \}
\end{gathered}
\end{equation}

\section {Buckmaster Equation Cost-Function}
\label{appendix:Buckmaster}

The quantum analogue of the Buckmaster differential operator is $\hat{O}\ket{\tilde{f}} = (\frac{\hat{\partial ^2}}{\partial x^2}\hat{D}^3_{\tilde{f}} + \frac{\hat{\partial}}{\partial x}\hat{D}^2_{\tilde{f}})\ket{\tilde{f}}$. With the Forward Euler scheme, our cost-function is

\begin{equation}
    C = ||\ket{f} - (\mathds{1} + \tau\hat{O})\ket{\tilde{f}}||^2
\label{costfuncbuck}
\end{equation}

\noindent After expanding with the inner product, substituting $\ket{f} = \lambda_0 \hat{U}(\mathbf{\lambda})\ket{0} = \lambda_0\ket{\psi}$ and the Adder expressions for the derivatives, and simplifying, we get

\begin{equation}
\begin{gathered}
C = \lambda_0^2 + 2\lambda_0\operatorname{Re}\{-\tilde{\lambda}_0\braket{\psi|\tilde{\psi}} + \frac{2\cdot4^n\tilde{\lambda}_0^4\tau}{L^2}\braket{\psi|\hat{D}^3_{\tilde{\psi}}|\tilde{\psi}} \\ 
- \frac{4^n\tilde{\lambda}_0^4\tau}{L^2}(\braket{\psi|\hat{A}\hat{D}^3_{\tilde{\psi}}|\tilde{\psi}}+\braket{\psi|\hat{A}^\dag\hat{D}^3_{\tilde{\psi}}|\tilde{\psi}})
+ \frac{\alpha2^{n-1}\tilde{\lambda}_0^3\tau}{L}(-\braket{\psi|\hat{A}\hat{D}^2_{\tilde{\psi}}|\tilde{\psi}}+\braket{\psi|\hat{A}^\dag\hat{D}^2_{\tilde{\psi}}|\tilde{\psi}})\}
\end{gathered}
\end{equation}

\section{Deterministic Kardar--Parisi--Zhang Equation Cost-Function}
\label{appendix:KPZ}

To time evolve this equation we must first train an auxiliary state $\ket{\chi}$ to represent $\frac {\hat{\partial}}{\partial x} \ket{\tilde{f}}$. Our cost-function is thus

\begin{equation}
    C = ||\ket{\chi} - \frac{\hat{\partial}}{\partial x}\ket{\tilde{f}}||^2
\end{equation}

\noindent Letting $\ket{\chi} = \theta_0\hat{U}(\mathbf{\theta})\ket{0} = \theta_0\ket{\phi}$ and $\ket{\tilde{f}} = \tilde{\lambda}_0\hat{U}(\mathbf{\tilde{\lambda}})\ket{0} = \tilde{\lambda}_0\ket{\tilde{\psi}}$ and substituting the spatial derivatives, we get.

\begin{equation}
    C_\chi = \theta_0^2 -\frac{2^n}{L}\tilde{\lambda}_0\theta_0\operatorname{Re}\{\braket{\phi|\hat{A}|\tilde{\psi}}-\braket{\phi|\hat{A}^\dag|\tilde{\psi}}\}
\end{equation}

\noindent Now, we can construct our quantum operator $\hat{O}\ket{\tilde{f}} = \alpha \frac{\hat{\partial} ^2}{\partial x^2}\ket{\tilde{f}} + \beta\hat{D}_\chi\ket{\chi}$. Letting $\ket{f} = \lambda_0 \hat{U}(\mathbf{\lambda})\ket{0} = \lambda_0\ket{\psi}$ and substituting into Eq.~\ref{costfuncbuck}, we get

\begin{equation}
\begin{gathered}
    C_f = \lambda_0^2 - 2\lambda_0\operatorname{Re}\{\tilde{\lambda}_0(1 - \frac{2\alpha\tau4^n}{L^2})\braket{\psi|\tilde{\psi}}\\
    +\tilde{\lambda}_0\alpha\tau\frac{4^n}{L^2}(\braket{\psi|\hat{A}|\tilde{\psi}}+\braket{\psi|\hat{A}^\dag|\tilde{\psi}})
    +\theta_0^2\tau\beta\braket{\psi|\hat{D}_\phi|\phi}\}
\end{gathered}
\end{equation}

\section{Hardware Demonstrations Circuit Diagrams}
\label{appendix:Circuits}

\begin{figure}[h]
\begin{tikzpicture}
\node at (-1.7, 1.6)[scale=1] {\textbf{(a)} $\braket{\psi|\hat{A}|\psi}$};
\node at (-2, 0)[scale=1]{
    \begin{quantikz}[column sep=0.15cm, row sep={1cm,between origins}]
    \lstick[wires=1]{$\ket{0}$} & \gate{H} & \ctrl{1} & \ctrl{2} & \gate{H} & \meter{} \\
    \lstick[wires=2]{$\ket{0}^{\otimes 2}$} & \gate[wires=2]{SO(4)(\lambda_1, \lambda_2)} & \targ{} & \qw & \qw & \qw &\qw \\
    \qw & \qw & \ctrl{-1} & \targ{} & \qw & \qw & \qw
  \end{quantikz}};
  
 \node at (5, 1.6)[scale=1] {\textbf{(b)} $\braket{\psi|\hat{D}_{\phi}|\psi}$};
 \node at (5, 0)[scale=1]{
    \begin{quantikz}[column sep=0.15cm, row sep={1cm,between origins}]
    \lstick[wires=1]{$\ket{0}$} & \gate{H} & \ctrl{2} & \ctrl{1} & \gate{H} & \meter{} \\
    \lstick[wires=1]{$\ket{0}$} & \gate[wires=1]{R_y(\lambda)} & \qw & \targ{} & \qw & \qw & \qw \\
    \lstick[wires=1]{$\ket{0}$} & \qw & \gate[wires=1]{R_y(\theta)} & \ctrl{-1} & \qw & \qw & \qw & \qw
  \end{quantikz}};

\end{tikzpicture}
\caption{Circuit Diagrams for Fig. \ref{fig:hardware}. Both expectation values are calculated using the Hadamard test. In \textbf{(a)}, the 2 qubit controlled Adder is implemented by the Toffoli and CNOT. In \textbf{(b)}, the bottom ancilla qubit is used to implement the controlled Diagonal gate.}
\end{figure}
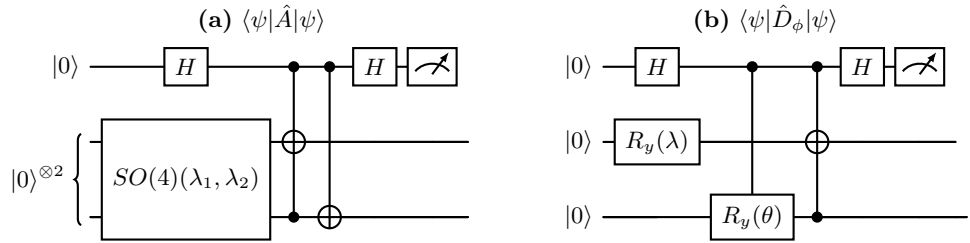

\end{document}